\def\be{\begin{equation}}
\def\ee{\end{equation}}
\def\ba{\begin{eqnarray}}
\def\ea{\end{eqnarray}}
\def\nn{\nonumber}

\def\ellmax{\ell_{\rm max}}
\def\zmax{z_{\rm max}}
\def\n{{\widehat{\bf n}}}
\def\bigoh{{\mathcal O}}
\def\hphi{{\widehat\phi}}

\def\hC{\widehat C}
\def\phips{\phi_{\rm ps}}
\def\ha{\widehat a}

\def\hE{\widehat E}
\def\del{({\rm delensed})}
\def\DeltaP{\Delta_P}

\newcommand{\threej}[6]{\left(
                           \begin{array}{ccc}
        \! #1\! & #2\!  & #3\!  \\
        \! #4\! & #5\!  & #6\!
                           \end{array}
                   \right)}

\documentclass[12pt,letterpaper]{article}

\usepackage{color}
\usepackage{amsmath}
\usepackage{amsfonts}
\usepackage{amssymb}
\usepackage{graphicx}
\usepackage[numbers,sort&compress]{natbib}

\usepackage[hang,small,bf]{caption}

\DeclareRobustCommand{\SkipTocEntry}[4]{}

\def\be{\begin{equation}}
\def\ee{\end{equation}}
\def\ba{\begin{eqnarray}}
\def\ea{\end{eqnarray}}

\begin{document}

\vspace{5mm} \vspace{0.5cm}

\begin{center}

{\Large CMBPol Mission Concept Study: Gravitational Lensing}
\\[1.0cm]

{Kendrick M. Smith$^{\dagger{\rm 1}}$, 
Asantha Cooray$^{\rm 2}$,
Sudeep Das$^{\rm 3,4}$,
Olivier Dor\'e$^{\rm 5}$,
Duncan Hanson$^{\rm 1}$,
Chris Hirata$^{\rm 6}$,
Manoj Kaplinghat$^{\rm 2}$,
Brian Keating$^{\rm 7}$,
Marilena LoVerde$^{\rm 8,9}$,
Nathan Miller$^{\rm 7}$,
Gra\c{c}a Rocha$^{\rm 6}$,
Meir Shimon$^{\rm 7}$,
and Oliver Zahn$^{\rm 10,11}$}
\\[0.5cm]

\end{center}

\vspace{2cm} \hrule \vspace{0.3cm}
{\small  \noindent \textbf{Abstract}} \\[0.3cm]
\noindent Gravitational lensing of the cosmic microwave background by large-scale structure in the late universe is both a
source of cosmological information and a potential contaminant of primordial gravity waves.
Because lensing imprints growth of structure in the late universe on the CMB, measurements of CMB lensing
will constrain parameters to which the CMB would not otherwise be sensitive, such as neutrino mass.

In CMB polarization, gravitational lensing is the largest guaranteed source of B-mode (or curl-like) polarization.
Future CMB polarization experiments with sufficient sensitivity to measure B-modes on small angular scales ($\ell\sim 1000$)
can measure lensing with better sensitivity, and on different scales, than could be achieved by measuring CMB
temperature alone.
If the instrumental noise is sufficiently small ($\lesssim 5$ $\mu$K-arcmin), the gravitational lensing contribution
to the large-scale B-mode will be the limiting source of contamination when constraining a stochastic background of gravity
waves in the early universe, one of the most exciting prospects for future CMB polarization experiments.
High-sensitivity measurements of small-scale B-modes can reduce this contamination through a lens reconstruction technique
that separates the lensing and primordial contributions to the B-mode on large scales.

A fundamental design decision for a future CMB polarization experiment such as CMBpol is whether to have coarse angular resolution
so that only the large-scale B-mode (and the large-scale E-mode from reionization) is measured, or high resolution to additionally
measure CMB lensing.
The purpose of this white paper is to evaluate the science case for CMB lensing in polarization:
constraints on cosmological parameters, increased sensitivity to the gravity wave B-mode via lens reconstruction,
expected level of contamination from non-CMB foregrounds, and required control of beam systematics.

 \vspace{0.5cm}  \hrule
\def\thefootnote{\arabic{footnote}}
\setcounter{footnote}{0}

\vspace{1.0cm}

\vfill \noindent
$^\dagger$ {\footnotesize {\tt kmsmith@ast.cam.ac.uk}}\\
\newpage

\begin{center}

{\small \textit{$^{\rm 1}$ Institute of Astronomy, University of Cambridge, Cambridge, CB3 0HA, UK}}  \\
{\small \textit{$^{\rm 2}$ Department of Physics and Astronomy, University of California, Irvine, CA 92697-4575}} \\
{\small \textit{$^{\rm 3}$ Department of Physics, Jadwin Hall, Princeton University, Princeton, NJ, 08544}} \\
{\small \textit{$^{\rm 4}$ Department of Astrophysical Sciences, Peyton Hall, Princeton University, Princeton, NJ 08544 }} \\
{\small \textit{$^{\rm 5}$ CITA, University of Toronto, 60 St George Street, Toronto, ON, M5S 3H8, Canada }} \\
{\small \textit{$^{\rm 6}$ California Institute of Technology, Pasadena, CA 91125, USA}} \\
{\small \textit{$^{\rm 7}$ Center for Astrophysics and Space Sciences, University of California,}} \\
{\small \textit{9500 Gilman Drive, La Jolla, CA, 92093-0424}}  \\
{\small \textit{$^{\rm 8}$ Institute for Strings, Cosmology and Astro-particle Physics (ISCAP)}} \\
{\small \textit{$^{\rm 9}$ Department of Physics, Columbia University, New York, NY 10027}} \\
{\small \textit{$^{\rm 10}$ Berkeley Center for Cosmological Physics, Department of Physics,}} \\
{\small \textit{University of California, Berkeley, CA 94720, USA}} \\
{\small \textit{$^{\rm 11}$ Lawrence Berkeley National Labs, University of California, Berkeley, CA 94720, USA}}

\end{center}

\newpage
\tableofcontents

\newpage
\section{Gravitational lensing and CMB polarization}

\subsection{Introduction}

Much of the progress in
cosmology in the last two decades has been due to the well understood
physics underlying the CMB anisotropy. The CMB promises to remain a
gold mine for precision cosmology, and two new frontiers lie
ahead. The first one is the primary purpose of this report, that is the
polarized component that offers the prospects of detecting primordial
gravitational waves and constraining recombination physics. Second,
large scale structures between the last scattering surface and us alters the primary CMB
anisotropy, through gravitational lensing (for a recent review of the
theory see \cite{Lewis:2006fu}). Other effects like the scattering off
hot electrons in large scale structure (the Sunyaev-Zel'dovich effects), and through
redshifting during the traverse of time-dependent potential
fluctuations (the ISW effect) are relevant for temperature and
will be mostly ignored here. In this section, we will show how
those two frontiers actually merge when looking at CMB
polarization at sub-degree angular scales. We will present how
gravitational lensing of the polarized CMB constitutes a unique
cosmological probe and the conceptual and practical challenges that arise.
The large scale density fluctuations in the universe induce random
deflections in the direction of the CMB photons as they propagate from
the last scattering surface to us. The displacement angle is related
to the projected surface density or, equivalently, the projected
gravitational potential. This effect can be rewritten as a remapping
of the primordial unlensed CMB field the following way:
\ba
T(\n) & = & T(\n + \nabla \phi(\n))   \nonumber\\
(Q\pm iU)(\n) & = & (Q\pm iU)(\n + \nabla \phi(\n))   \label{eq:deflection}
\ea
where the deflection angle $\nabla \phi$ is expressed in terms of the gravitational potential as
\be
\phi(\n) = -2 \int_0^{z_{\rm rec}} \frac{dz}{H(z)} \Psi(z,D(z)\n)
\left( \frac{D(z_{\rm rec})-D(z)}{D(z_{\rm rec})D(z)} \right)\ ,  \label{eq:line_of_sight}
\ee
where $D(z)$ denotes the comoving distance to redshift $z$ in the assumed flat cosmology and $\Psi(z,{\bf x})$ is the zero-shear gravitational potential.
In the Limber approximation, the power spectrum of $\phi$ is given by:
\be
C_\ell^{\phi\phi} = \frac{8\pi^2}{\ell^3} \int_0^{z_{\rm rec}}
\frac{dz}{H(z)} D(z) 
\left( \frac{D(z_{\rm rec})-D(z)}{D(z_{\rm rec})D(z)} \right)^2\ .
P_\Psi(z,k=\ell/D(z)) 
\label{eq:clphiphi}
\ee

Since the structures as described by the gravitational potential $\Psi$ are
not very correlated on large scales, the gravitational lensing effect is
only relevant at small angular scales in the CMB. This fact has made
CMB lensing observationally challenging so far. Nevertheless, using
cross correlation between WMAP data and other tracers of large scale
structures to increase the signal to noise, a detection of
gravitational lensing in the CMB temperature has been achieved with
marginal significance, i.e. around 3$\sigma$
\cite{Hirata:2004rp,Smith:2007rg,Hirata:2008cb}. A direct detection in temperature
is expected to be achieved soon with high significance thanks to
on-going high angular resolution temperature surveys 
(e.g. ACT \cite{Kosowsky:2006na}, SPT \cite{Ruhl:2004kv}, Planck \cite{planck}).

Promisingly, it was realized that the lensing of the CMB is more
significant in polarization than in temperature \cite{Zaldarriaga:1998ar}. This stems from
the fact that the lensing effects on the CMB can be qualitatively
understood as a smearing of the CMB acoustic peaks in the angular
power spectrum. Since the CMB polarization has sharper acoustic peaks
than temperature, the gravitational lensing effect is more significant
in polarization than in temperature by approximately a factor of two. But the
instrumental sensitivity required to detect the lensing effect in
polarization is nevertheless higher than for temperature because of the weak degree of
polarization of the CMB in the first place.

\begin{figure}[t]
\begin{center}
\includegraphics[angle=90,width=0.50\textwidth]{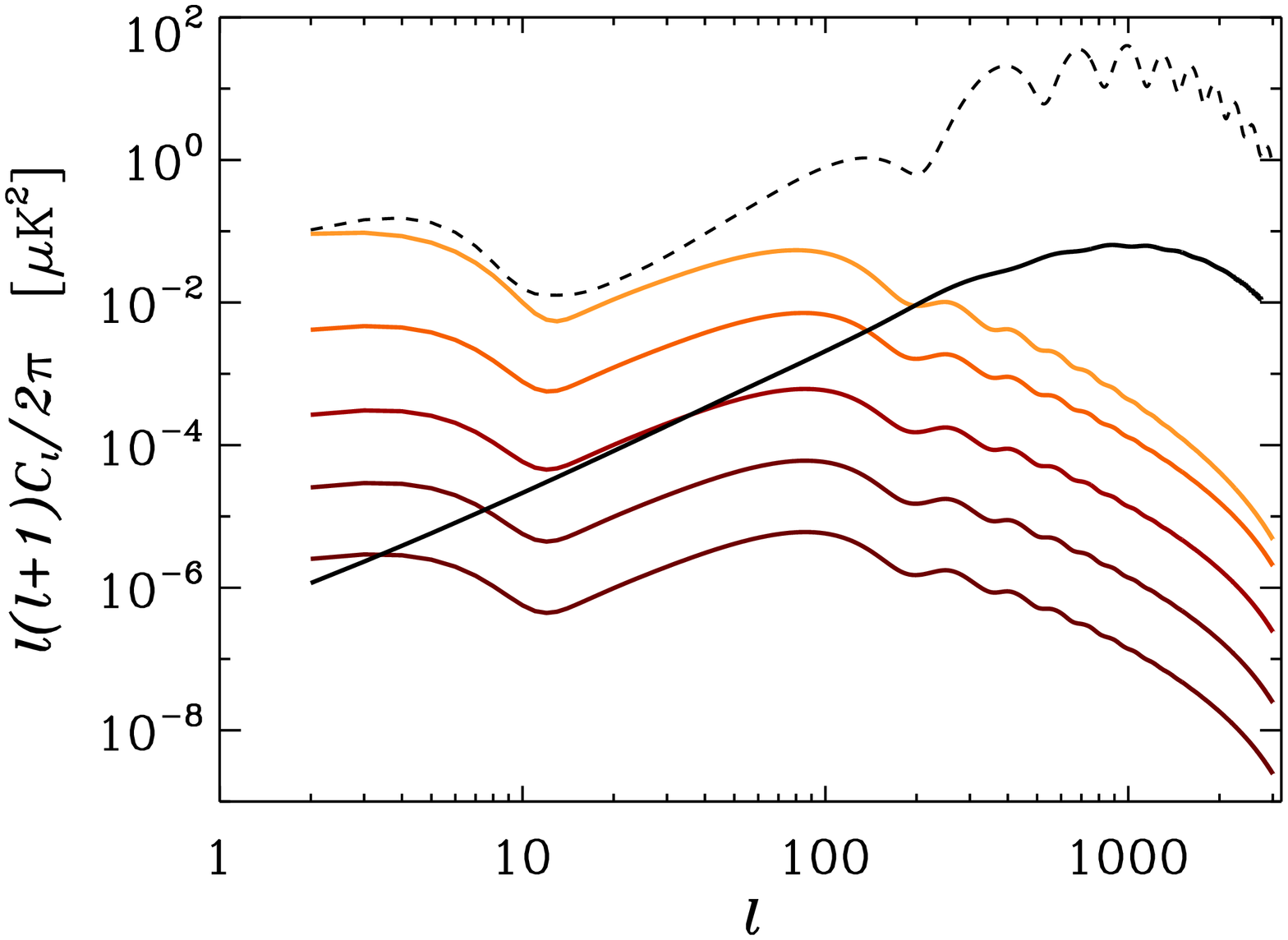}
\includegraphics[angle=0,width=0.46\textwidth]{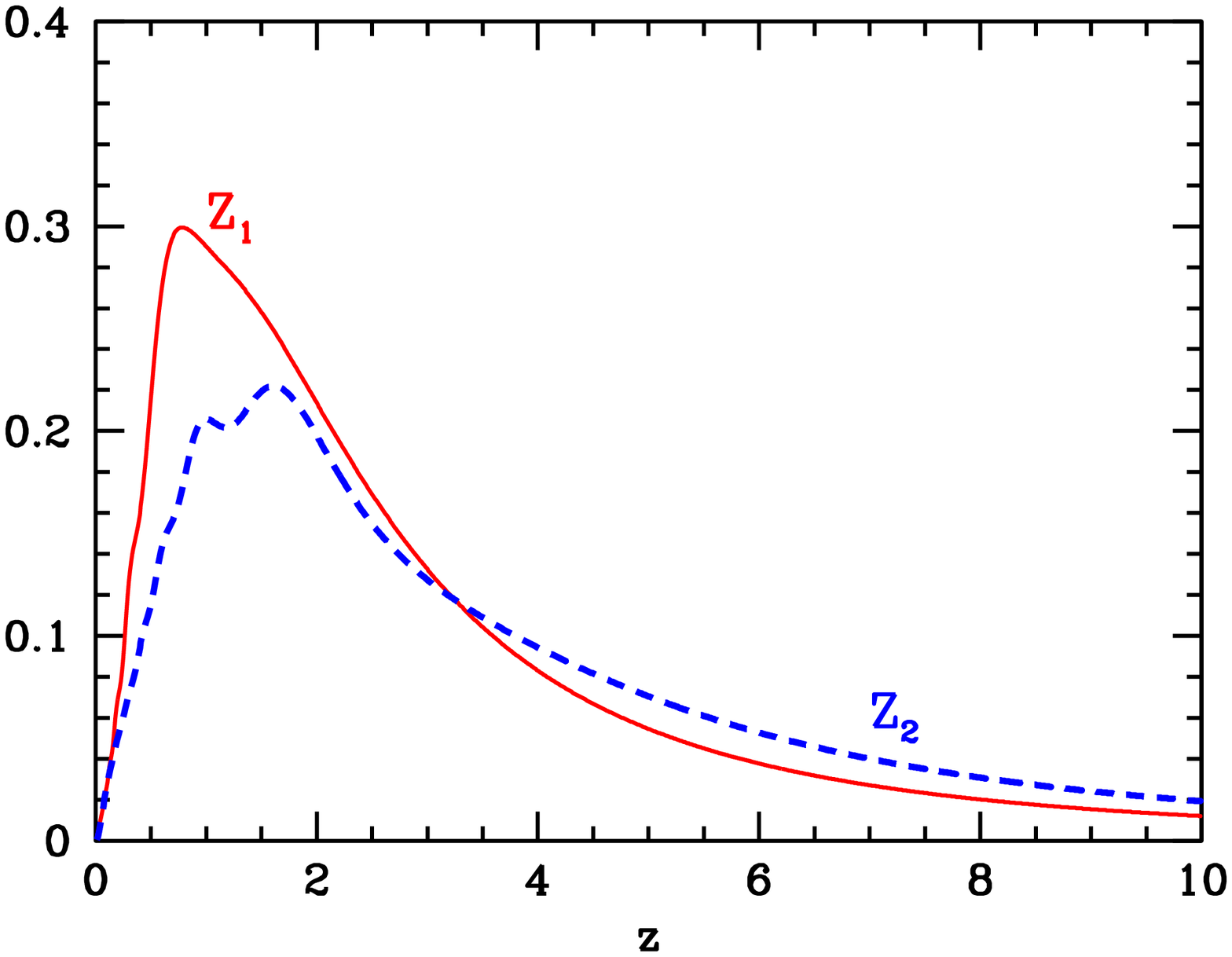}
\end{center}
\caption{{\it Left panel:} Signal angular power spectrum for the E (dashed line) and B
 (solid lines) modes. The black solid dashed line corresponds to the
 lensing induced B modes for all the models considered. The light to
 dark red colored curves correspond to different $r$ values, namely
 0.43, 0.1, 0.01 and 0.001. The cosmological parameters used for this
 plot correspond to the WMAP5 $\Lambda$CDM+$r$ best fit model
 \cite{Komatsu:2008hk}. Note that $r=0.43$ corresponds to the 95\% upper limit on $r$ using
 this data-set. Obviously, for any allowed value of $r$, the lensing signal
 will dominate for $\ell\ge 200$. {\it Right panel:} Redshift
 dependence of the two principal components ($Z_1$ and $Z_2$ respectively) of the lensing potential
 angular power spectrum defined in Eq.~(\ref{eq:clphiphi}) (from
 \cite{Smith:2006nk}). These curves illustrate the CMB
 polarization lensing sensitivity to moderate redshifts, i.e. up to $z\simeq 5$.
}\label{fig:clbb_ref}
\end{figure}

However, the lensing of the polarized CMB presents several interesting
features. First, as seen in Eq.~(\ref{eq:deflection}),
gravitational lensing lensing does not mix Q and U, it will
nevertheless result in a mixing of the E and B modes because the transformation from (Q,U)
to (E,B) is non-local \cite{Zaldarriaga:1996xe,Kamionkowski:1996ks,Zaldarriaga:1998ar}. In particular, E mode power will be
transferred into B modes, generating in this way the largest guaranteed
B-mode signal. This particular signal is totally independent from the existence of
primordial B modes, i.e. the existence of tensor modes in the early
universe as illustrated in the left panel of Fig.~\ref{fig:clbb_ref}. Since for realistic
values of $r$, this B mode lensing signal is likely to dominate over
the primordial one at sub-degree scales, it might limit our quest from
primordial B mode \cite{Knox:2002pe,Kesden:2003cc,Seljak:2003pn} if
not properly taken care of. The procedure of cleaning the lensing
signal or ``delensing'' the B modes will be made explicit below.

Although a contaminant when trying to measure $r$, CMB polarization
by itself contains unique cosmological information. Being sensitive to both the geometry of the
universe and the growth of structure at moderate redshift
(z$\lesssim$5) as illustrated in the right panel of Fig.~\ref{fig:clbb_ref}, the CMB
lensing breaks the angular diameter distance degeneracy in the CMB. It
gives us a unique handle on the universe expansion history between
recombination and moderate redshifts that is a rare probe of early dark
energy. It provides access to the deepest two dimensional mass maps
possible, thus anchoring tomographic studies of the evolution of dark
energy at lower redshifts. CMB polarization lensing also provides a unique
opportunity to map the distribution of matter on large scales and
high redshifts where density fluctuations are still in the linear
regime and are thus robust cosmological probes. On smaller scales,
CMB lens reconstruction can directly probe halo mass profiles,
without any need to calibrate cluster masses against other observables
such as SZ temperature \cite{Seljak:1999zn,Dodelson:2004as,Vale:2004rh,Maturi:2004zj,Holder:2004rp,Lewis:2005fq,Hu:2007bt,Hu:2007jh,Yoo:2008bf}.
 Furthermore, since the
lensing B-modes allow for an order of magnitude extension to smaller
scales of the lensing potential as compared to temperature lensing, it
is uniquely sensitive to parameters that affect structure formation in
the late universe, such as neutrino masses \cite{Lesgourgues:2005yv,Smith:2006nk}.

It must be said however that holding these promises is observationally
demanding. Gravitational lensing of the polarized CMB is a small scale
manifestation of the very large scale properties of the intervening
mass distribution. It therefore requires both high angular resolution
($\lesssim$ 10 arcmin) and wide-field surveys ($\gtrsim$ square degrees) to be
exploited. This comes of course at an additional cost  and complexity
for a satellite mission that must be quantitatively weighted against
the scientific returns. This section aims at providing the science
elements relevant to this debate.

\subsection{Lens reconstruction and delensing}
\label{ssec:intro_lens_reconstruction}

The most powerful techniques for extracting the gravitational lensing signal from the CMB are based on the idea of ``lens reconstruction'',
in which the deflection operation in Eq.~(\ref{eq:deflection}) is inverted statistically: starting from the lensed (observed) CMB, one defines
an estimator $\hphi_{\ell m}$ for the lens potential (which is not directly observable)
\cite{Bernardeau:1996aa,Bernardeau:1998mw,Zaldarriaga:1998te,Benabed:2000jt,Guzik:2000ju,Hu:2001fa,Hu:2001tn,Hu:2001kj,
Knox:2002pe,Kesden:2002ku,Hirata:2002jy,Kesden:2003cc,Hirata:2003ka,Seljak:2003pn}.

To understand intuitively how this is possible, imagine that both the lensed E-mode and B-mode have been measured with high signal-to-noise.
Because there is no unlensed B-mode, the deflection operation (Eq.~(\ref{eq:deflection})) converts two unobserved fields (the unlensed E-mode and the
lens potential) into two observed fields (the lensed E-mode and B-mode).  Inverting the deflection operation, to recover the unobserved fields
from the observed ones, is possible (at least at the level of counting degrees of freedom) because it amounts to solving for two free fields
given the values of two observed fields.\footnote{This intuitive description fails to capture some qualitative features of lens reconstruction;
for example that lens reconstruction can be done (at lower signal-to-noise) from CMB temperature alone, or that joint estimation of a gradient
and curl mode in the deflection angles is possible.  However, it does give a simple intuitive interpretation of the polarization estimator in
the high signal-to-noise limit.}

On a technical level, lens reconstruction is possible because the B-mode generated by gravitational lensing is highly correlated to the E-mode,
with a correlation whose ``shape'' depends on the realization of the lens potential $\phi$.
In a fixed realization of the lens potential, the EB two-point function is of the form (see App.~\ref{app:lens_reconstruction}):
\be
\left\langle a_{\ell_1 m_1}^E a_{\ell_2 m_2}^B \right\rangle_{\rm CMB} = \sum_{\ell m} \Gamma^{(\phi)EB}_{\ell_1\ell_2\ell}\threej{\ell_1}{\ell_2}{\ell}{m_1}{m_2}{m} \phi_{\ell m}^*  \label{eq:gamma1}
\ee
where we have used the notation $\langle\cdot\rangle_{\rm CMB}$ to emphasize that the expectation value is taken over CMB realizations
in a fixed realization of $\phi$.  (Notation in Eq.~(\ref{eq:gamma1}) and elsewhere in the paper follows Dvorkin \& Smith, to appear \cite{DvorkinSmith}.)

By summing (with minimum variance weighting) over two-point terms in the CMB which average to a given mode $\phi_{\ell m}$ of the lensing potential, 
we can write down an estimator $\hphi_{\ell m}$ for the mode:
\ba
\hphi_{\ell m} &=& N_\ell^{\phi\phi} \sum_{\ell_1m_1\ell_2m_2} \Gamma^{EB}_{\ell_1\ell_2\ell} \threej{\ell_1}{\ell_2}{\ell}{m_1}{m_2}{m} a_{\ell_1m_1}^{E*} a_{\ell_2m_2}^{B*}  \label{eq:hphi_def} \\
N_\ell^{\phi\phi} &=& \left[ \frac{1}{2\ell+1} \sum_{\ell_1\ell_2} \frac{|\Gamma^{EB}_{\ell_1\ell_2\ell}|^2}{(C_{\ell_1}^{EE}+N_{\ell_1}^{EE})(C_{\ell_2}^{BB}+N_{\ell_2}^{BB})} \right]^{-1}  \label{eq:nl_def}
\ea
This estimator is unbiased, in the sense that:
\be
\left\langle \hphi_{\ell m} \right\rangle_{\rm CMB} = \phi_{\ell m}
\ee
and its covariance is given\footnote{Eq.~(\ref{eq:hphi_covariance}) is actually an approximation; it includes most, but not all, of the contractions in the CMB
four-point function.  The additional terms can be interpreted as a change of normalization in the power spectrum estimator and removed using an
iterative method \cite{Cooray:2002py}.} by:
\be
\left\langle \hphi_{\ell m}^* \hphi_{\ell' m'} \right\rangle = (C_\ell^{\phi\phi} + N_\ell^{\phi\phi}) \delta_{\ell\ell'} \delta_{mm'}  \label{eq:hphi_covariance}
\ee
We therefore interpret $\hphi_{\ell m}$ as a noisy reconstruction of the lens potential $\phi$, with noise power spectrum given by the quantity $N_\ell^{\phi\phi}$
defined in Eq.~(\ref{eq:nl_def}).
(Note that the expectation value $\langle\cdot\rangle$ in Eq.~(\ref{eq:hphi_covariance}) is taken over realizations of the CMB and lens potential.)

\begin{figure}
\begin{center}
\includegraphics[width=4in]{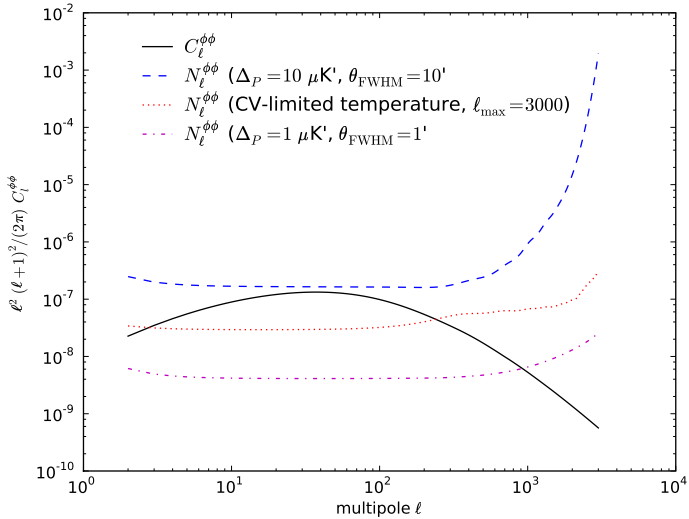}
\end{center}
\caption{Signal power spectrum $C_\ell^{\phi\phi}$ for the CMB lens potential, and reconstruction noise power spectra
for low-noise (1 $\mu$K-arcmin noise, 1 arcmin beam) and high-noise (10 $\mu$K-arcmin noise, 10 arcmin beam) polarization measurements,
and for temperature measurements which are cosmic variance limited to $\ellmax=3000$.}
\label{fig:nlphi}
\end{figure}

In Fig.~\ref{fig:nlphi}, we show some example signal and noise power spectra for the EB quadratic estimator, for low-noise (1 $\mu$K-arcmin noise, 1 arcmin beam)
and high-noise (10 $\mu$K-arcmin noise, 10 arcmin beam) polarization measurements.  As the instrumental sensitivity varies over this range, the lens reconstruction
goes from having signal-to-noise $\lesssim 1$ on all angular scales, to being a high signal-to-noise reconstruction out to sub-degree scales ($\ell\lesssim 1000$).
For comparison, we also show a noise power spectrum for the TT quadratic estimator, assuming cosmic variance limited observations out to $\ellmax=3000$.\footnote{We
have chosen an $\ellmax$ cutoff here, rather than assuming a noise level and beam size in temperature, because futuristic lens reconstruction measurements from CMB
temperature are more likely to be limited by foregrounds on small scales than by instrumental noise \cite{Amblard:2004ih}.}
It is seen that the signal-to-noise of the reconstruction drops sharply for $\ell\gtrsim 200$ even for cosmic variance limited observations, i.e. reconstructing the
smallest scales in $\phi$ require measuring polarization and cannot be done from CMB temperature alone \cite{Hu:2001kj}.

In this report, we will concentrate on two applications of the lens reconstruction estimator $\hphi_{\ell m}$.
First, the power spectrum estimator
\be
\hC_\ell^{\phi\phi} = \left( \frac{1}{2\ell+1} \sum_{m=-\ell}^\ell \hphi_{\ell m}^* \hphi_{\ell m} \right) - N_\ell   \label{eq:hcl_def}
\ee
is useful as a direct probe of large-scale structure, and used to constrain quantities such as neutrino mass to which the primary CMB is not sensitive.
(Note that we define $\hC_\ell^{\phi\phi}$ in Eq.~(\ref{eq:hcl_def}) with the noise bias term from Eq.~(\ref{eq:hphi_covariance}) subtracted.)
Such constraints could also be obtained from the B-mode power spectrum $C_\ell^{BB}$ \cite{Stompor:1998zj,Lewis:2005tp,Smith:2006nk}, 
but performing lens reconstruction allows more cosmological information
to be extracted from the lensing signal.
The overall signal-to-noise is higher,
and parameter degeneracies can be broken, in cases where two parameters are degenerate in the B-mode power spectrum but produce distinct effects on the
power spectrum $C_\ell^{\phi\phi}$

The second application of lens reconstruction that we will study in this report is ``delensing'', or reducing the level of the lensing B-mode as a contaminant
of the gravity wave signal from inflation.  On an intuitive level, delensing can be described as follows.  Suppose that the instrumental noise is sufficiently
low (and foregrounds and systematics sufficiently well-controlled) that the lensing B-mode on large scales is the dominant source of noise when estimating the
tensor-to-scalar ratio $(T/S)$.
In this low-noise regime, the large-scale B-mode $a_{\ell m}^B$ has been measured with high signal-to-noise, but is a sum of lensing and primordial
contributions, and sample variance of the lensing component dominates the uncertainty $\sigma(T/S)$.
If we have a reconstruction $\hphi$ of the lens potential, and we also have measurements of the E-mode on intermediate angular scales ($20 \lesssim\ell\lesssim 2000$),
then we can simply perform the deflection operation (Eq.~(\ref{eq:deflection})) to obtain a reconstruction $\ha^B_{\ell m}$ of the lensed B-mode on large scales.
We then ``delens'' the observed B-mode by subtracting this reconstruction ($a^B_{\ell m} \rightarrow a^B_{\ell m} - \ha^B_{\ell m}$), to obtain a new large-scale
B-mode in which the level of lensing power has been reduced, while preserving the primordial contribution.
An estimate of the tensor-to-scalar ratio which is based on this ``delensed'' B-mode will therefore have a smaller uncertainty $\sigma(T/S)$.
A more formal version of this delensing procedure, which incorporates noise using minimum-variance weighting, will be given later in this report
(\S\ref{sec:delensing}, App.~\ref{app:lens_reconstruction}).

\subsection{Foregrounds and systematics}
\label{ssec:intro_foregrounds_systematics}

Studies of lens reconstruction to date have mainly focused on the statistical errors that can be obtained assuming that
the microwave sky consists of a Gaussian primary CMB, lensed by a potential $\phi$.
In reality, there are astrophysical sources of radiation at microwave frequencies: either diffuse Galactic foregrounds (synchrotron
radiation, free-free emission, dust emission from either vibrational or rotational modes of the grains) or extragalactic signals (point sources,
thermal/kinetic SZ).
These foreground signals are particularly worrying for lens reconstruction because they are not Gaussian fields, and lens reconstruction
can be interpreted as constraining lensing via its non-Gaussian signature in the CMB (e.g., the estimated power spectrum $\hC_\ell^{\phi\phi}$
can be viewed as a trispectrum estimator \cite{Zaldarriaga:2000ud,Hu:2001fa,Okamoto:2002ik,Cooray:2002py}).
At the time of this writing, foreground bias in lens reconstruction is largely unexplored territory (see however \cite{Amblard:2004ih} for some
results on temperature foregrounds).
In \S\ref{sec:foregrounds} we will argue that in polarization, the picture is relatively simple: extragalactic polarized point sources are expected
to generate the largest foreground bias.
We calculate the bias for a realistic model of the flux and redshift distribution of radio sources, and argue that foregrounds are not expected
to bias lens reconstruction significantly, for a wide range of noise levels and beam sizes, if the reconstruction is done using polarization.

Another practical concern for lens reconstruction from CMB polarization (or for any measurement which makes use of B-modes
in a critical way) is instrumental systematics, particulaly beam systematics \cite{Hu:2002vu,Rosset:2004jj,O'Dea:2006di,Shimon:2007au}.
Beam systematics can be classified into reducible (effects which are coupled to the scan strategy) and irreducible (effects which persist
for an ideal survey), and further subclassified into specific effects (e.g. differential pointing).
For each beam systematic, the bias to lens reconstruction can be computed using the formalism from \cite{Miller:2008zi}, and the instrumental
limit on the systematic effect (e.g. as measured from Jupiter maps) can be compared to the threshhold for producing a statistically significant
bias in cosmological parameters such as $m_\nu$ or $(T/S)$.
This provides a framework for studying systematics that will be presented in detail in \S\ref{sec:systematics}.

\subsection{Outline}

The outline of this White Paper is as follows.
In \S\ref{sec:parameter_forecasts}, we  study CMB lensing as a source of cosmological information, presenting forecasts
in cases where lensing adds qualitatively new cosmological information (compared to what could be obtained using the unlensed
CMB alone): neutrino mass, the dark energy of state $w$, and mean curvature.
In \S\ref{sec:delensing}, we consider the lensing B-mode as a contaminant to the gravity wave signal on large scales, and
forecast prospects for delensing, or reducing the level of contamination using measurements of the small-scale lensing potential
to reconstruct the lens potential and the lensing B-mode.
We also consider ``external'' delensing using datasets other than small-scale polarization: either small-scale CMB {\em temperature}
(\S\ref{ssec:temperature_delensing}) or large-scale structure (\S\ref{ssec:lss_delensing}), but conclude that these approaches are not
promising.
In \S\ref{sec:foregrounds} we consider the impact of foregrounds.  We argue that polarized extragalactic point sources are likely to
be the dominant foreground component for lens reconstruction, and forecast the level of contamination due using realistic modeling
of radio sources.
Finally, in \S\ref{sec:systematics}, we study beam systematics and compute tolerance levels for quantities such as differential pointing
or beamwidth, guided by the criterion that the systematic error on cosmological parameters such as $(T/S)$ or $m_\nu$ should be a small
fraction of the statistical error.

\newpage
\section{Parameter forecasts}
\label{sec:parameter_forecasts}

In this section, we consider CMB lensing as a source of information on cosmological parameters.
As described in \S\ref{ssec:intro_lens_reconstruction}, the lens reconstruction estimator $\hphi_{\ell m}$ allows us to extract a noisy 
measurement of the CMB lensing potential $\phi_{\ell m}$ from high-resolution observations of the CMB.
In effect, we can observe an extra field: the resulting measurement of $C_\ell^{\phi\phi}$ can be folded into a cosmological parameter
analysis along with the direct measurement of the CMB power spectra $C_\ell^{TT}, C_\ell^{TE}, C_\ell^{EE}$.

\subsection{Cosmological information in the unlensed CMB}

How does the cosmological information in $C_\ell^{\phi\phi}$ compare to the information contained in the CMB power spectra $C_\ell^{TT}, C_\ell^{TE}, C_\ell^{EE}$?
To answer this question, let us temporarily ignore CMB lensing, and ask what cosmological information is contained in the unlensed CMB power spectra.
The qualitative picture we will give in this section is explored in much greater detail in e.g. 
\cite{Hu:1996qs,Zaldarriaga:1997ch,Metcalf:1997ih,Stompor:1998zj,Hu:2000ti,Hu:2001bc}.

The shape of the CMB power spectra is directly sensitive to parameters which affect the physics of the evolving plasma in the early universe, 
such as the baryon density $\Omega_b h^2$, the matter density $\Omega_m h^2$, and the shape of the primoridal power spectrum
(parameterized through a spectral index $n_s$ or perhaps additional parameters describing ``running'' of the spectral index with scale).
Additionally, the overall amplitude of the power spectra is proportional to $A_s e^{-2\tau}$, where $A_s$ denotes the amplitude of the initial fluctuations and 
$\tau$ denotes the optical depth to recombination.
This introduces a degeneracy between $A_s$ and $\tau$ that can be broken ``internally'' to the CMB by measuring the E-mode reionization bump on large scales, 
which is sensitive to $\tau$ alone.
(For more discussion of reionization and CMB polarization, we refer the reader to the companion white paper \cite{cmbpol_reionization}.)

The unlensed CMB contains another parameter degeneracy, the ``angular diameter distance degeneracy'', which arises when one attempts to constrain ``late universe''
parameters which mainly affect distances and growth after recombination.
In this section, we will consider the following late universe parameters: the dark energy density $\Omega_\Lambda$, dark energy equation of state $w$, curvature $\Omega_K$,
and neutrino mass $(\sum m_\nu)$.
Such parameters only affect the CMB through the angular scale of the acoustic peaks $\ell_a$, which is a ratio of two distances:
\be
\ell_a = \pi \frac{D_*}{s_*}
\ee
where $D_*$ is the angular diameter distance to recombination and $s_*$ is the sound horizon, or total distance that a sound wave can travel between the big bang and recombination.
In a parameter space containing $N$ late universe parameters, one combination of the parameters will be well-constrained by the unlensed CMB (via the angular diameter distance $D_*$),
leaving a near-perfect $(N-1)$-fold degeneracy between the others.

\begin{figure}
\begin{center}
\includegraphics[width=5in]{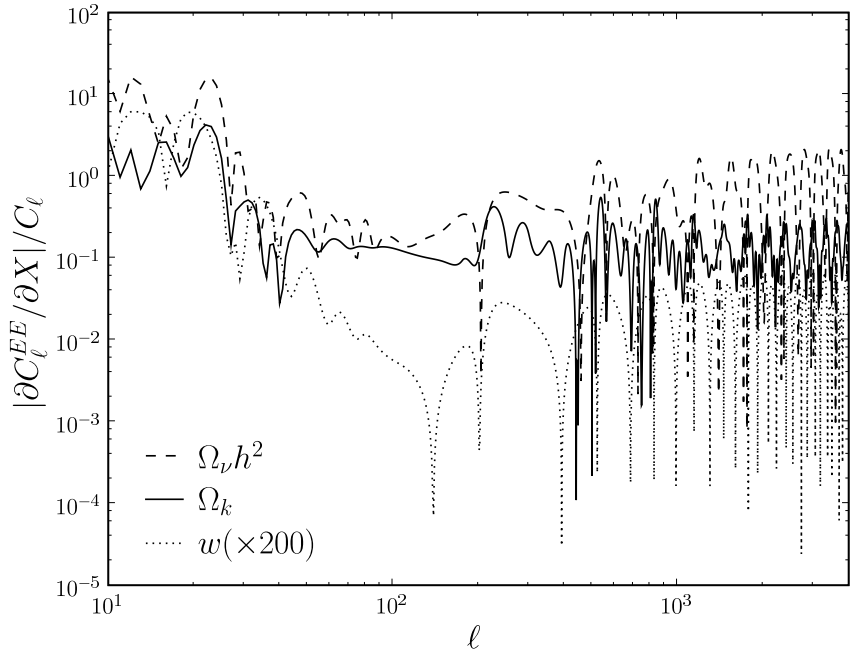}
\end{center}
\caption{Derivatives of the {\em unlensed} $C_\ell^{EE}$ power spectrum with respect to the late universe parameters $\{\Omega_\nu h^2, w, \Omega_K\}$ along the angular diameter distance
degeneracy, showing that the power spectra remain constant to a good approximation.}
\label{fig:unlensed_degeneracy}
\end{figure}

Another way of describing the angular diameter distance degeneracy is that if we vary any of the parameters $\{\Omega_\nu h^2, w, \Omega_K\}$, adjusting the dark energy density
$\Omega_\Lambda$ so that the angular diameter distance to recombination $D_*$ remains fixed, then the unlensed CMB power spectra remain fixed to an excellent approximation.
This is directly illustrated in Fig.~\ref{fig:unlensed_degeneracy}, where we plot the derivative of the unlensed power spectrum $C_\ell^{EE}$ with respect to each of these three parameters along
the degeneracy surface $D_*=$constant.  
It is seen that for cosmologically interesting step sizes in these parameters (say $\Delta\Omega_\nu h^2=0.01$, $\Delta w=0.2$, $\Delta\Omega_k=0.01$) the fractional change in $C_\ell^{EE}$ is
very small and the unlensed CMB is essentially unchanged.

\subsection{Cosmological information from CMB lensing}

\begin{figure}
\begin{center}
\includegraphics[width=5in]{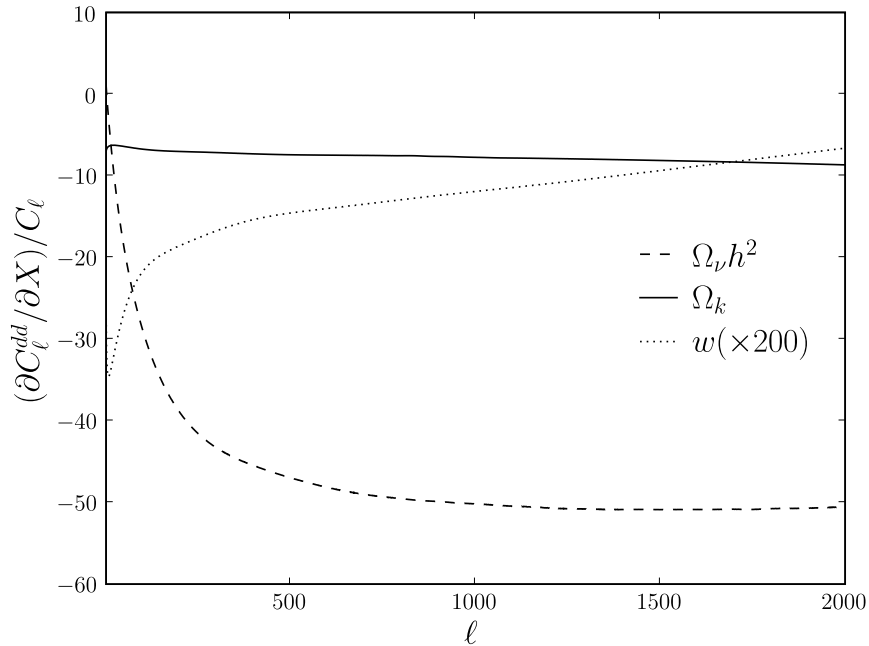}
\end{center}
\caption{Derivatives of $C_\ell^{\phi\phi}$ with respect to the same late universe parameters as in Fig.~\ref{fig:unlensed_degeneracy}, showing a large change in the power
spectrum as the parameters are varied: the angular diameter distance degeneracy is broken by lensing.}
\label{fig:lensed_degeneracy}
\end{figure}

Lens reconstruction presents the possibility of breaking the angular diameter distance degeneracy in the unlensed CMB, by measuring the power spectrum $C_\ell^{\phi\phi}$ of
the lens potential.  This power spectrum can be written as a line-of-sight integral which includes both geometric distances and the power spectrum of the evolving potential
(Eq.~(\ref{eq:line_of_sight})), so it depends on both distances and growth and is generally sensitive to late universe parameters such as $\{ \Omega_\nu h^2, w, \Omega_K \}$.
This can be seen explicitly in Fig.~\ref{fig:lensed_degeneracy}, where we show the derivative of the power spectrum $C_\ell^{\phi\phi}$ with respect to each of the three parameters,
taking the derivative along the degeneracy surface $D_*=$constant as in Fig.~\ref{fig:unlensed_degeneracy}.
Comparing the two figures, it is seen that measurements of the CMB lens potential do break the angular diameter distance degeneracy, allowing each of these three parameters to be
constrained from the CMB alone.

Constraining late universe parameters through lensing is a future application of CMB experiments which measure the small-scale modes,
and for experiments which measure small-scale polarization in particular.
As remarked in the introduction, in the limit of low noise and high resolution, CMB polarization experiments can ultimately reconstruct the modes of $\phi_{\ell m}$ with
high signal-to-noise across a wider range of angular scales $(\ell\lesssim 1000)$ than are accessible using CMB temperature alone.
In the next few subsections, we will present forecasts for parameter constraints from CMB lensing, using a Fisher matrix formalism described in detail in App.~\ref{app:fisher}.

We include unlensed temperature and polarization power spectra (TT, EE, TE)
in our analysis and include the lensing information through the deflection
angle power spectrum. 
We do not use the lensed power spectra to avoid the complication of
the correlation in their errors between different $\ell$ values and
with the error in $C_\ell^{\phi\phi}$. Using the lensed spectra and neglecting
these correlations could lead to overly optimistic 
forecasts \cite{Hu:2001fb}.  A previous study \cite{Kaplinghat:2003bh} found
that using lensed spectra instead of the unlensed ones (plus $\phi_{\ell m}$
power spectrum) shrunk the expected errors on $w$ and $m_\nu$ for
their version of CMBpol by about 40\% and 30\% respectively.  

We now consider neutrino mass, dark energy and curvature in turn and forecast
the sensitivity of {\em CMB alone} to constrain these late-universe parameters. 

\subsection{Neutrino mass}
\label{ssec:neutrino_mass}

Neutrinos are a part of the standard model of particle physics and it is now
known from neutrino oscillation experiments that neutrinos are not massless and
that the three known mass-eigenstates are not fully degenerate.  The atmospheric
\cite{Hosaka:2006zd,Sanchez:2003rb,Ambrosio:2004ig} and solar neutrino
experiments \cite{Fukuda:2002pe,Smy:2003jf,Aharmim:2005gt} as well as
experiments with man-made neutrino beams
\cite{Ahn:2006zza,Abe:2008ee} have measured two mass-square differences to
be close to $8 \times 10^{-5}$ eV$^2$ and $3 \times 10^{-3}$ eV$^2$.
This implies that there must be at least one active neutrino with a mass
greater than about 0.05 eV. Fortuitously, both CMB lensing and cosmic shear
experiments can get to this level of sensitivity
\cite{Kaplinghat:2003bh,Abazajian:2002ck}. We note that the lensing experiments
are sensitive to the sum of the neutrino masses and it is possible that the
neutrinos are highly degenerate with a sum of masses close to or larger than
0.15 eV.

\noindent
{\em Limits on neutrino mass.} 
The neutrino oscillation experiments measure the mass-squared differences,
but not the sum of the neutrino masses. The most stringent laboratory upper
bound on absolute neutrino mass comes from tritium beta decay
end-point experiments \cite{Bonn:2002rz} which limit the electron neutrino mass
to $\lesssim 2$ eV. This could improve by an order of magnitude in the future
with the KATRIN experiment \cite{Bonn:2008zz}.  There are other proposed
experiments that plan to get to similar sensitivity and detection limits
by searching for neutrinoless double beta decay \cite{Zdesenko:2001ee}.  A Dirac
mass would elude this search, but theoretical prejudice favors and the
see-saw mechanism requires Majorana masses. Like the CMB and galaxy
shear observations, these future tritium end-point and neutrinoless double
beta decay experiments will be extremely challenging. 

The current large scale structure surveys (2dFGRS, SDSS) and WMAP
together already provide powerful constraints on neutrino mass. We know that
the sum of the active neutrino masses is less than about 0.7
eV \cite{Komatsu:2008hk}. The sum of the active neutrino masses, $m_\nu$,
is related to their energy density $\Omega_\nu h^2 \approx m_\nu/(94 \mbox{eV})$
assuming thermally populated neutrinos. As mentioned earlier, at the
lower end, atmospheric neutrino oscillations constrain the mass of at least
one active neutrino to be larger than about 0.05 eV. This window from 0.05 eV
to about 1 eV can be probed with both laboratory experiments and
cosmological observations. 

A change in $m_\nu$ gives rise to many effects. First, it changes the expansion
rate of the universe. At last scattering, this leads to a change in the sound
horizon and damping length (of the photon-baryon fluid). The change in the
sound horizon shifts the position of the peaks and troughs in the
anisotropy spectrum while the change in the damping length (relative to the
sound horizon) changes its amplitude. Second, the presence of a relativistic
or semi-relativistic species has an effect on the CMB even after last
scattering because it causes the gravitational potential to change (decay)
with time. The photons traversing these potential wells red-shift or
blue-shift, and this enhances the amplitude of the anisotropy spectrum. The
above two effects are however degenerate with other parameters, most notably
the matter density. 

There is, however, a third effect that is distinct -- on small scales,
the presence of a massive neutrino damps the growth of structure. The
net suppression of the power spectrum of density fluctuations is scale
dependent and the relevant length scale is the Jeans length for
neutrinos \cite{Bond:1983hb,Ma:1996za,Hu:1997vi} which decreases with time as
the neutrino thermal speed decreases. This suppression of growth is
ameliorated on scales larger than the Jeans length at matter--radiation
equality, where the neutrinos cluster like cold dark matter. Neutrinos
never cluster on scales smaller than the Jeans length today. The net result is
no effect on large scales and a suppression of power on small scales. This
effect can be used to put constraints on the neutrino mass using the
observed galaxy power spectrum combined with CMB observations
\cite{Hu:1997mj}. Eisenstein et al. \cite{Eisenstein:1998hr} predicted that
the primary CMB spectrum from the Planck satellite can measure neutrino mass
with an error of 0.26 eV. 

The alteration of the gravitational potential at late times changes
the gravitational lensing of CMB photons as they traverse these
potentials. Including the gravitational lensing effect, the Planck
error forecast improves to about 0.15 eV
\cite{Kaplinghat:2003bh,Perotto:2006rj}, with more ambitious experiments capable
of probing down to 0.05 eV level \cite{Kaplinghat:2003bh,Lesgourgues:2005yv}.
Tomographic observations of the galaxy shear due to gravitational lensing can a
achieve similar sensitivity in $m_\nu$ \cite{Abazajian:2002ck}. The physics in
both cases is the same: gravitational lensing. However the observations and the
associated systematics are very different. Complementary techniques are
valuable since these measurements will be very challenging.

\begin{figure}
\centerline{
\includegraphics[width=3in]{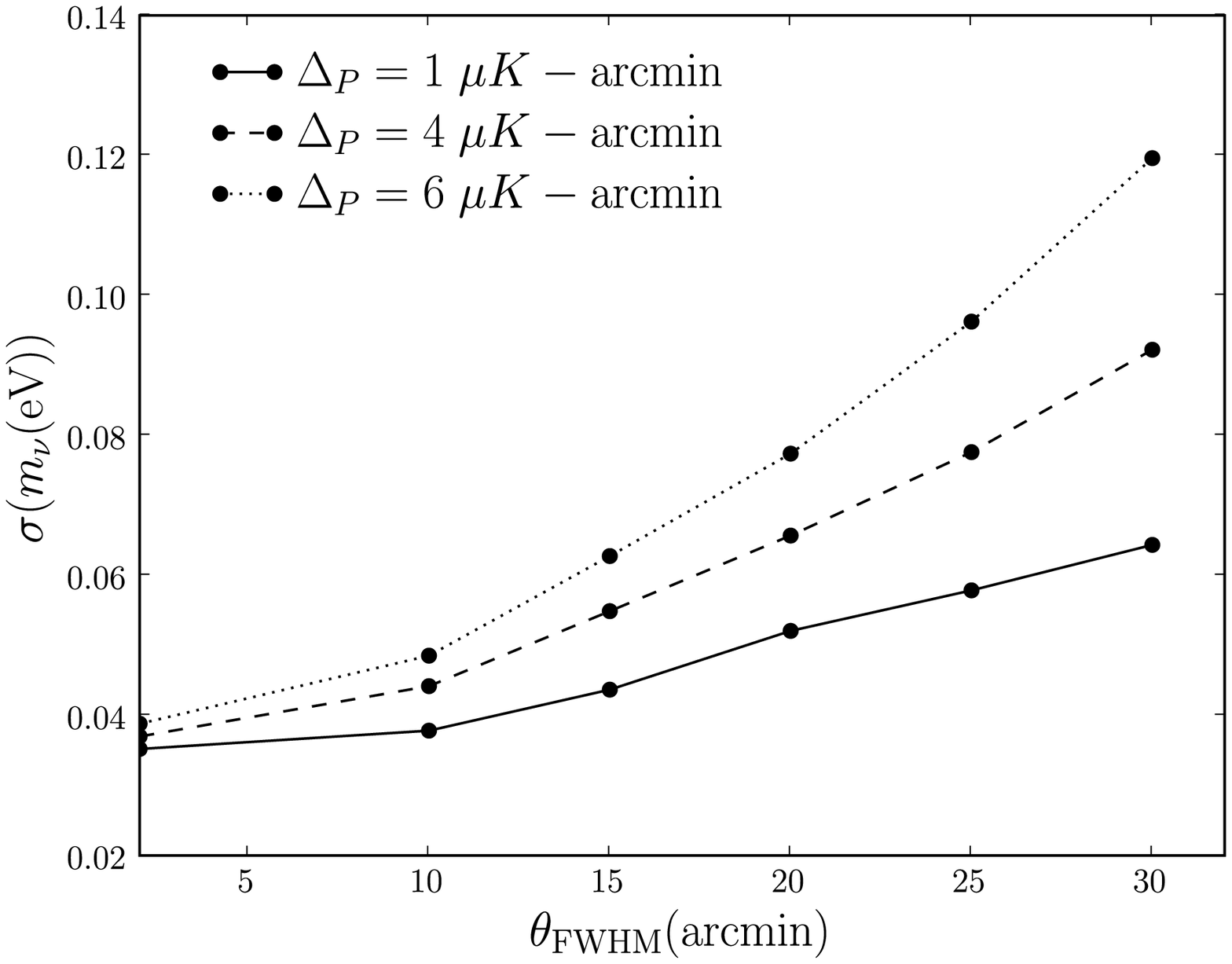}
\includegraphics[width=3in]{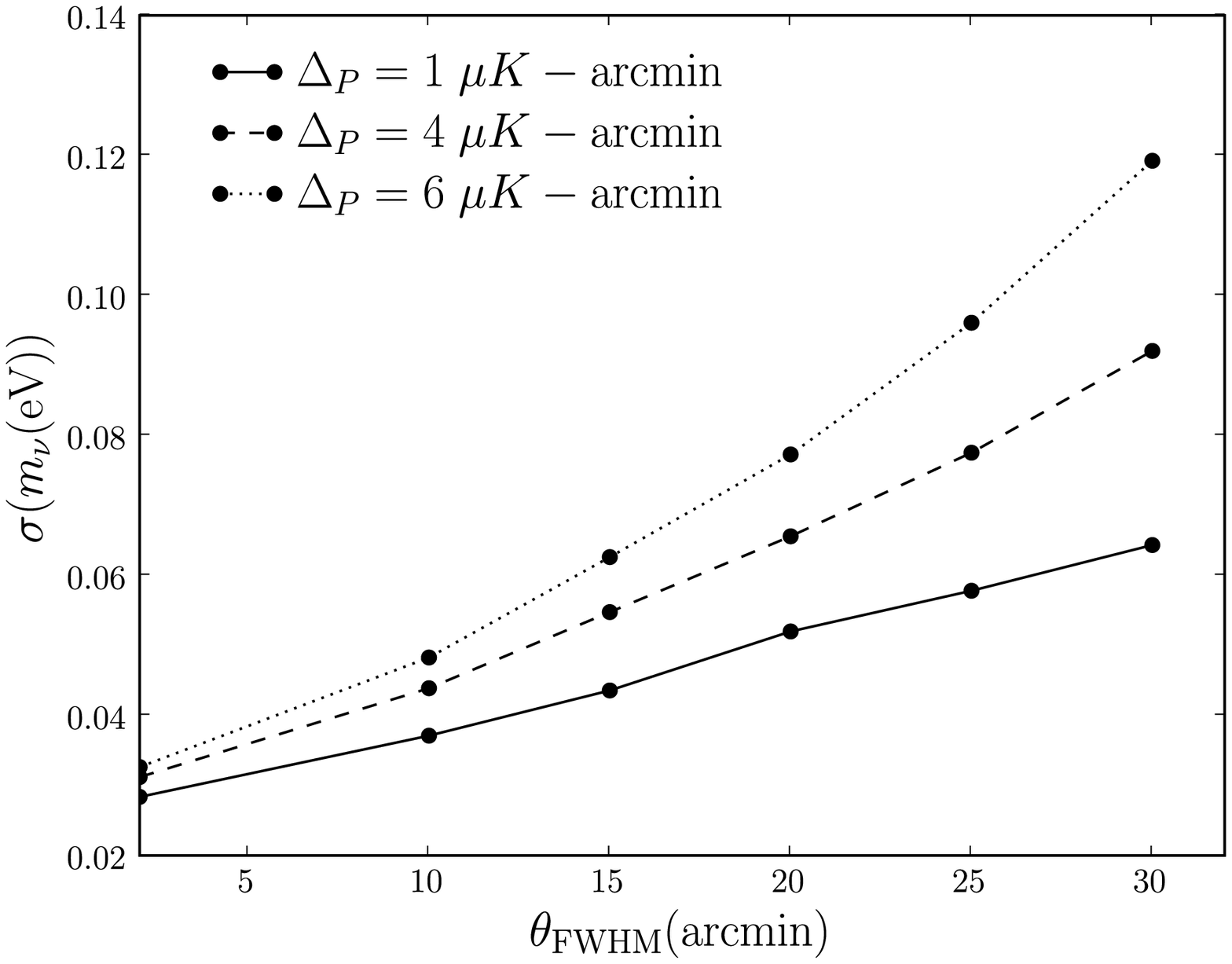}
}
\caption{Uncertainty $\sigma(m_\nu)$ on the neutrino mass as a function of beam size and noise level for $\ellmax=2000$ (left panel) or $\ellmax=4000$ (right panel)
using CMB lens reconstruction, assuming fixed $w,\Omega_K$.}
\label{fig:neutrino_mass_forecast}
\end{figure}

In order to forecast constraints on the neutrino mass, we marginalize over
the ``early universe'' parameters $\{ \Omega_bh^2, \Omega_ch^2,
\Omega_\Lambda, Y_{\rm He}, \tau, A, n_s \}$ and $\Omega_\Lambda$, but
assume that the parameters $w,\Omega_K$ are fixed.  (We will consider
joint constraints among $\{\Omega_\nu h^2, w, \Omega_K\}$
in \S\ref{ssec:joint_forecasts}.)  The result is shown in Fig.~\ref{fig:neutrino_mass_forecast}.
We find that a satellite mission can constrain $\sum_\nu m_\nu$, where the sum is taken over neutrino
species $\nu$, at the 0.03--0.12 eV level depending on the noise level $\Delta_P$, beam
width $\theta_{\rm FWHM}$ and maximum CMB multipole $\ellmax$ used in the lens reconstruction.

\noindent
{\em Summary of neutrino mass.}
The signature of a 0.1 eV neutrino in the unlensed CMB anisotropy spectra is
small and such small masses are only detectable through their effect on
lensing, which comes through their influence on the gravitational
potential. Future experiments like Planck will be able to statistically
detect the lensing effect and thus measure or put upper limits on the
neutrino mass. The expected 1-$\sigma$ error on $m_\nu$ from Planck is 0.15
eV, while CMBpol could get down to the 0.05 eV level to measure the neutrino
mass.

\subsection{Dark energy}

Dark energy affects lensing in two distinct ways. First, the presence of dark
energy implies faster expansion and hence a decrease in the overall growth
rate. Second, dark energy can cluster appreciably if the equation of state is
not identically 1. The second effect cannot be modeled unless we have
a microphysical description. Two simple approaches that are
common in the literature are to (1) model dark energy as one or more scalar
field(s) with possibly non-canonical kinetic terms (e.g.,
\cite{Wetterich:1987fm,Peebles:1987ek,Caldwell:1997ii,Wang:1999fa,
ArmendarizPicon:2000dh}) and (2) model dark energy as a perfect fluid with
a parameterized sound speed (e.g., \cite{Hu:1998tk,Erickson:2001bq}). We will
use the first approach with canonical kinetic terms in the following analysis.
These models are collectively called {\em quintessence}. The effect of the
dark energy density on the growth is easy to calculate for small scales where
the clustering is irrelevant. On larger scales, where dark energy
clusters appreciably it is no longer possible to factor the matter
density fluctuations in Fourier space into a part that depends on time
and another that depends on wave number. This was investigated in detail for
a constant equation of state by Ma et al. \cite{Ma:1999dwa}, who found that
dark energy clusters on scales $k \lesssim k_Q \equiv 2V_{,QQ}^{-1/2}$ where
dark energy has been modeled as a scalar field $Q$ with effective
mass $V_{,QQ}^{1/2}$ which is typically not much larger than ${\cal O}(H_0)$.
The clustering of dark energy boosts the metric perturbations and hence
lensing and thus CMB lensing offers a way to constrain dark energy properties
\cite{Hu:2001fb}. The primary effect is an overall suppression of the growth
factor except on large scales. 

\begin{figure}
\begin{center}
\centerline{
\includegraphics[width=3in]{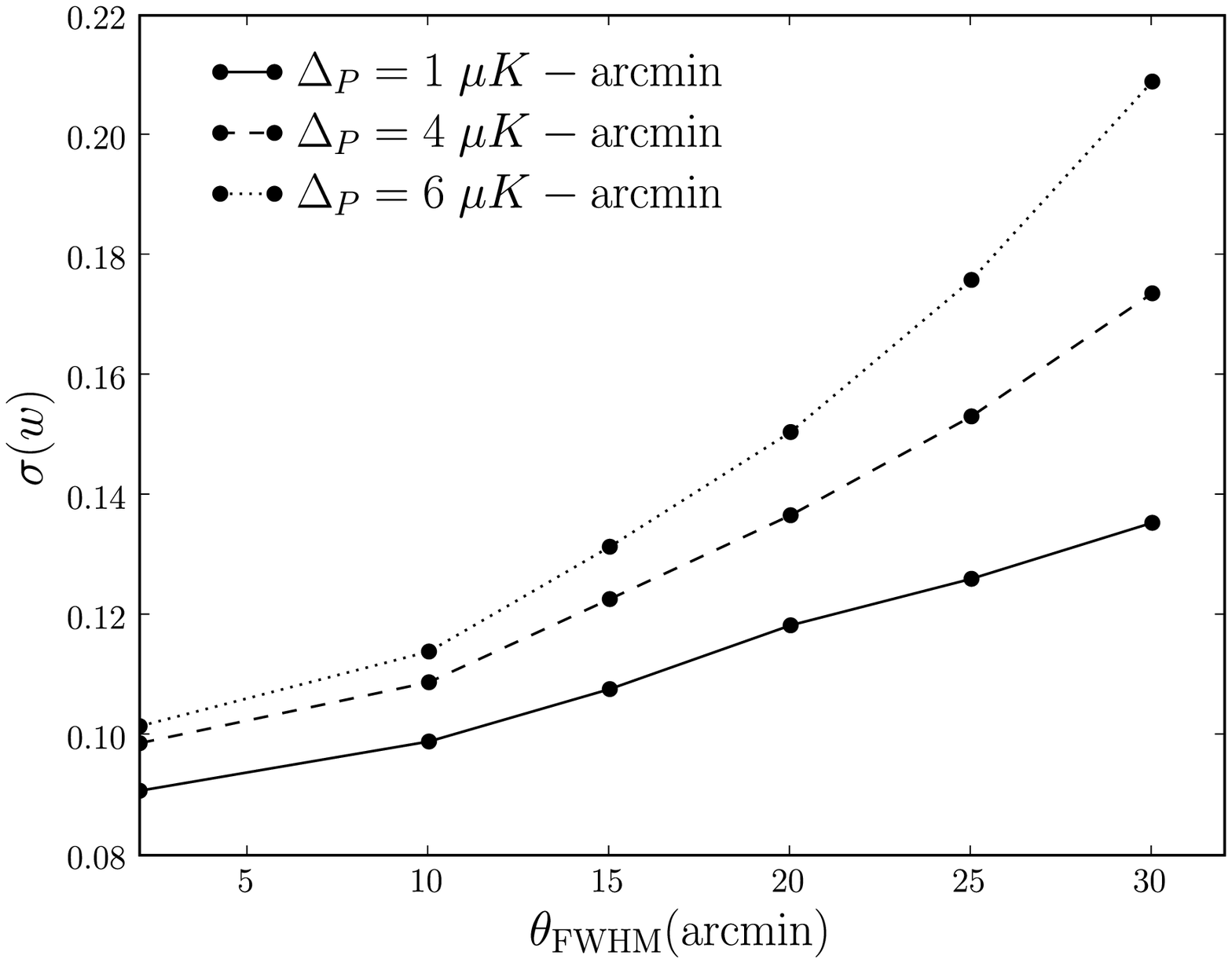}
\includegraphics[width=3in]{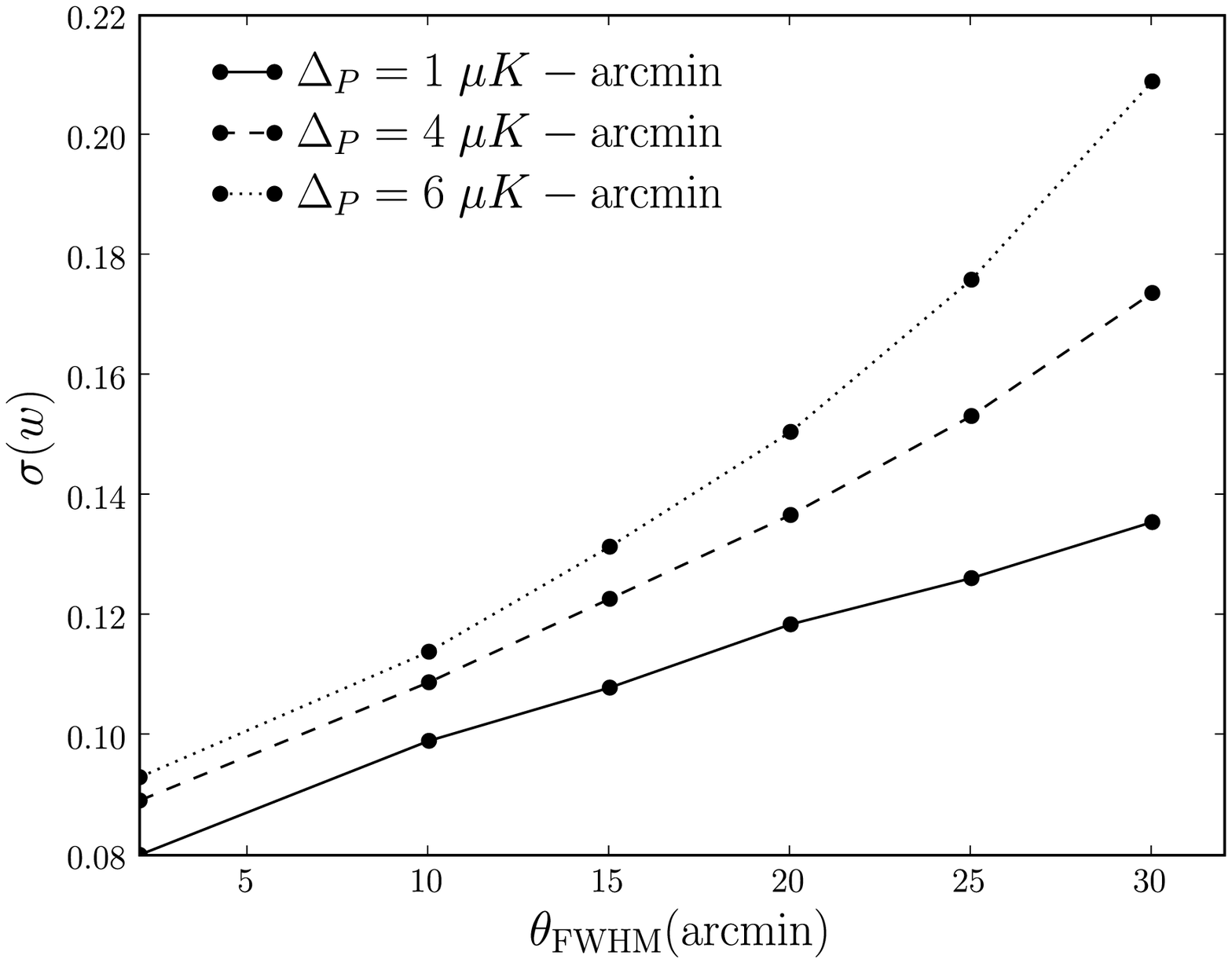}
}
\end{center}
\caption{Uncertainty $\sigma(w)$ on the dark energy equation of state as a function of beam size and noise level for $\ellmax=2000$ (left panel) or $\ellmax=4000$ (right panel)
using CMB lens reconstruction, assuming fixed $\Omega_\nu h^2,\Omega_K$.}
\label{fig:w_forecast}
\end{figure}

The CMB lensing window function is fairly broad in redshift--space. A
downside of this is that CMB lensing will never be
competitive with SNIa observations or proposed cosmic shear and
BAO experiments as far as measuring the equation of state of dark energy is
concerned. However, the virtue of CMB lensing is that it is an independent
alternative probe of the acceleration of the universe. 
CMBpol can measure $w$ to a precision of 0.08--0.2 depending on the noise
level and beam size (Fig.~\ref{fig:w_forecast}).

The sensitivity to a broad range of redshifts also 
implies that CMB lensing is a unique probe of dark energy (more
generally clustering) at $z > 2$. Note that if $w_X$ is
demonstrably different from -1, then dark energy must cluster on (at least)
large (1000 Mpc) scales and then the clustering properties of dark energy, say
parameterized in terms of its sound speed, might then be measurable
(e.g., \cite{Bean:2003fb}). 

The broadness of the window function also implies that the CMB is sensitive
to dark energy properties at high redshift. For the simplest quintessence
models, the contribution of dark energy at high redshifts is negligible.
However, there is no good reason to take these models as more than
possible examples. An important question is then that of the contribution of
dark energy to the expansion of the universe and growth of structure in the
early universe. There are many motivations to consider such extensions. Among
the most striking concerns is that of the timing coincidence: why is the
vacuum energy density (or scalar field potential) precisely small enough to
just begin dominating the energy density of the universe when the universe
grew to its present size? In this context, models with early dark energy
are arguably more natural
\cite{Albrecht:1999rm,ArmendarizPicon:2000dh,Dodelson:2001fq,Griest:2002cu,
Corasaniti:2002vg,Doran:2006kp} than simple quintessence where dark energy
emerges as a low-redshift phenomenon. Dark energy could also be an effect that
arises on horizon scales such as an infra-red modification to GR (e.g.,
\cite{Dvali:2000hr}). Note that the dark energy does not have to cause the
expansion of the universe to accelerate at early times.

We do not make predictions here for early dark energy because the predictions
depend on the models used
\cite{Dodelson:2001fq,Skordis:2000dz,Doran:2007ep}. If the low-redshift
dark energy equation of state parameters are constrained by other experiments
(such as SNIa), then CMB lensing should be able to measure the (average)
high-redshift equation of state at least as well as the constant equation of
state $w$ considered here and thus provide an unique window into the expansion
history and growth of structure at high redshifts.

\subsection{Curvature and joint constraints}
\label{ssec:joint_forecasts}

\begin{figure}
\centerline{
\includegraphics[width=3in]{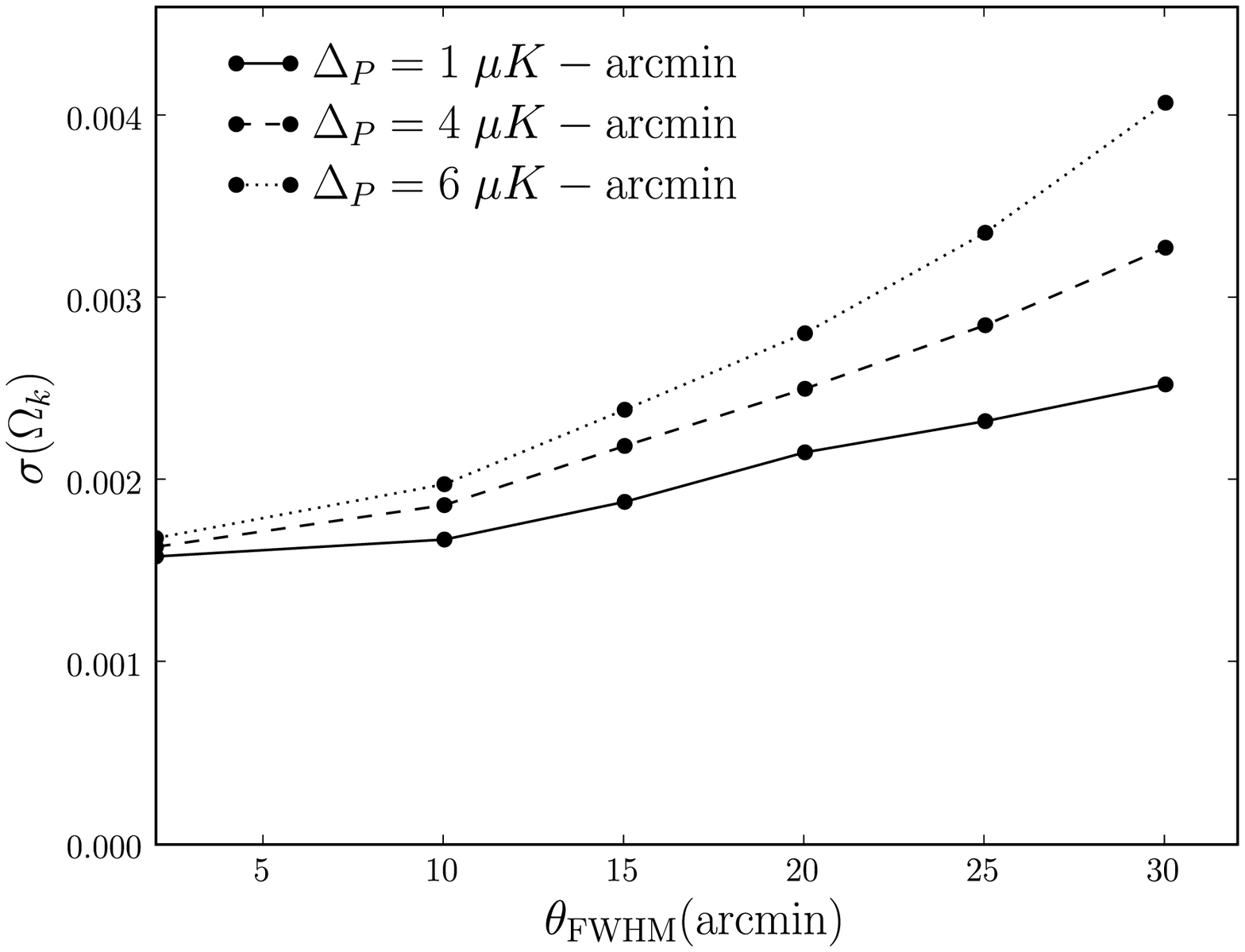}
\includegraphics[width=3in]{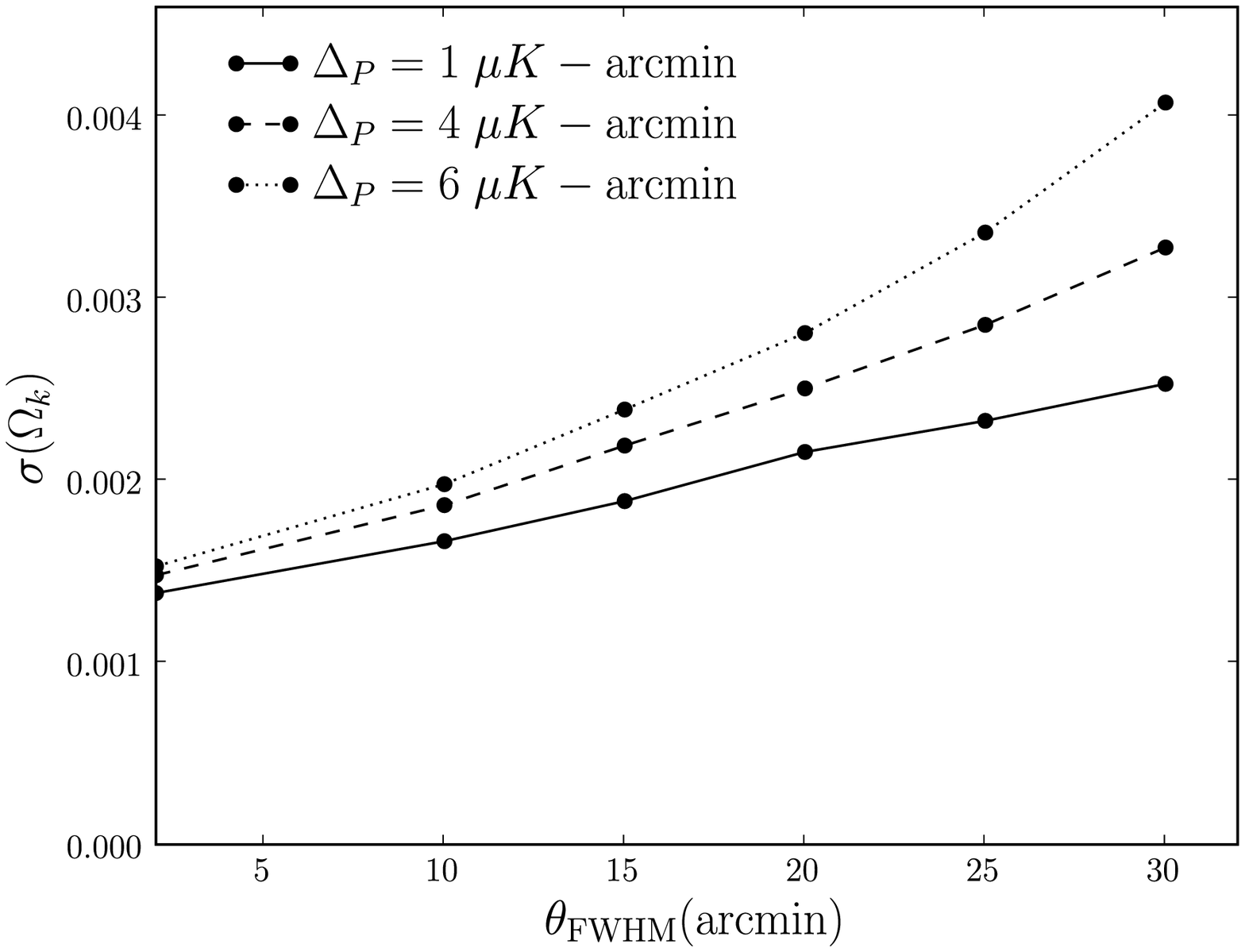}
}
\caption{Uncertainty $\sigma(\Omega_k)$ on mean curvature as a function of beam size and noise level for $\ellmax=2000$ (left panel) or $\ellmax=4000$ (right panel)
using CMB lens reconstruction, assuming fixed $m_\nu$ and $w$.}
\label{fig:curvature_forecast}
\end{figure}

As a further example of a parameter constraint from CMB lensing, we consider the mean curvature $\Omega_K$.  
(Historically, this was the first example of a new parameter constraint from breaking the angular diameter degeneracy
via CMB lensing \cite{Metcalf:1997ih}.)
The mean curvature is expected to be small in most inflationary cosmologies \cite{Guth:1981} but there are interesting
inflationary models with $\Omega_k = \bigoh(10^{-2})$; this is roughly the current $1\sigma$ upper limit from
combining CMB, BAO, and SN datasets \cite{cmbpol_inflation}.
In Fig.~\ref{fig:curvature_forecast} we show $1\sigma$ forecasts for the uncertainty $\sigma(\Omega_k)$, assuming that
$m_\nu$ and $w$ are fixed to fiducial values.
A polarization satellite can obtain $\sigma(\Omega_k) = (\mbox{few} \times 10^{-3})$ from the CMB alone.

Finally, we discuss joint constraints.  In Figs.~\ref{fig:neutrino_mass_forecast},~\ref{fig:w_forecast},~\ref{fig:curvature_forecast}
we have computed uncertainties on each of the three ``late universe'' parameters $\{ m_\nu, w, \Omega_k \}$ with the other two parameters
in this set fixed to fiducial values.
One can ask, in a parameter space in which all three late universe parameters are floating, can they be simultaneously constrained, or
are there degeneracies?
To quantify this, we compute the 3-by-3 correlation matrix between the late universe parameters in the Fisher formalism with all the early
universe parameters marginalized.  The result is
\be
\left( \begin{array}{ccc}
1 & 0.34 & -0.82 \\
0.34 & 1 & -0.63 \\
-0.82 & -0.63 & 1
\end{array} \right)  \label{eq:fisher33}
\ee
where the ordering of rows and columns is $m_\nu,w,\Omega_k$.
(This matrix was computed assuming $\Delta_P = 1.4$ $\mu$K-arcmin and $\theta_{\rm FWHM}=3$ arcmin.
The correlations are significantly different from zero but not so large that we would describe this parameter space as containing a degeneracy.
The strongest correlation is between curvature and the other parameters.
This is makes intuitive sense given the $C_\ell^{\phi\phi}$ derivatives shown in Fig.~\ref{fig:lensed_degeneracy}, 
where the fractional change in $C_\ell^{\phi\phi}$ with respect to curvature is approximately constant and highly
correlated with the derivative with respect to neutrino mass and $w$.

In conclusion, all-sky measurements of CMB polarization with high sensitivity and resolution can qualitatively
add information to the unlensed CMB: using lens reconstruction, the neutrino mass can be constrained to roughly
$\sigma(\sum m_\nu)=0.05$ eV, dark energy equation of state to roughly $\sigma(w)=0.15$, and mean curvature to
roughly $\sigma(\Omega_k)=2.5\times 10^{-3}$.
The precise values will depend on the noise level and beam as shown in Figs.~\ref{fig:neutrino_mass_forecast},~\ref{fig:w_forecast},~\ref{fig:curvature_forecast}.
Because the shape of the $C_\ell^{\phi\phi}$ power spectrum is reconstructed, and the shape dependence with respect to each of the three 
parameters $\{ \sum m_\nu, w, \Omega_k \}$ is different (Fig.~\ref{fig:lensed_degeneracy}), there are no degeneracies in this parameter space
although the correlations between parameters are significantly different from zero (Eq.~(\ref{eq:fisher33})).

\newpage
\section{Delensing the gravity wave B-mode}
\label{sec:delensing}

Perhaps the most exciting prospect for future generations of high-sensitivity CMB polarization experiments is constraining
the tensor-to-scalar ratio $(T/S)$, by measuring B-mode polarization on large scales.
In inflationary models, the value of $(T/S)$ is tied directly to the energy scale during inflation, so that measuring this
value opens a window on the physics that gave rise to the initial conditions of our universe (c.f. the companion white paper \cite{cmbpol_inflation}).
In a real experiment, the parameter uncertainty $\sigma(T/S)$ will receive contributions from detector noise, foreground
contamination, instrumental systematics, and contamination due to lensing B-modes.
In this section, we will consider the last of these contributions: under what circumstances is gravitational lensing the limiting
factor in measuring $(T/S)$, and what are the prospects for reducing the lensing contamination using delensing methods?
Many results from this section have been taken from \cite{delensing}, where more details will be presented.

It is easy to compare the contributions to $\sigma(T/S)$ from gravitational lensing and detector noise.
(The contribution from polarized foregrounds is studied in the companion white papers \cite{cmbpol_foreground_removal,cmbpol_foreground_science};
we will present some analysis of instrumental systematics in \S\ref{sec:systematics}.)
If we restrict attention to large angular scales ($\ell\lesssim 100$), then the lensing B-mode power spectrum $C_\ell^{BB}$ is constant to an 
excellent approximation, and the statistics of the lensed B-mode can be treated as Gaussian \cite{Smith:2004up,Smith:2005ue,Li:2006pu}.
Therefore lensing can be simply be thought of as an excess source of white noise.
The RMS amplitude $\sigma^B_{\rm lensing}$ of the lensing B-mode on large scales is $\approx 5$ $\mu$K-arcmin; if
the instrumental noise $\sigma_{\rm inst}$ is $\gtrsim \sigma^B_{\rm lensing}$, then the lensing contribution to $\sigma(T/S)$ is smaller than the noise contribution;
if $\sigma_{\rm inst} \lesssim \sigma^B_{\rm lensing}$, then lensing dominates.

For experiments with $\sigma_{\rm inst} \lesssim 5$ $\mu$K-arcmin, the only possibility for reducing the level of lensing contamination is to use 
``delensing'' techniques.  As described in \S\ref{ssec:intro_lens_reconstruction}, delensing can be performed whenever we have a noisy template $\hphi_{\ell m}$ for 
the CMB lens potential, and noisy measurements of the primary E-mode on intermediate scales.
We will consider several possibilities for the template $\hphi_{\ell m}$: it could either be obtained ``internally'' from CMB polarization on small
angular scales (\S\ref{ssec:polarization_delensing}), or ``externally'' from a different dataset, either small-scale CMB {\em temperature} (\S\ref{ssec:temperature_delensing}) 
or observations of large-scale structure (\S\ref{ssec:lss_delensing}).

In each of these cases, we will present forecasts for the parameter uncertainty $\sigma(T/S)$ with and without delensing.
Our forecasting methodology is presented in detail in App.~\ref{app:lens_reconstruction}, but let us note one key point here.
The effect of delensing is to change the equivalent white noise level of the large-scale B-mode from the value $\sigma^B_{\rm lensing}\approx 5$ 
$\mu$K-arcmin to some smaller value $\sigma^B_{\rm delensed} \le \sigma^B_{\rm lensed}$.\footnote{This statement is actually empiricial; in the forecasting
methodology from App.~\ref{app:lens_reconstruction}, we calculate a complete power spectrum $C_\ell^{B'B'}$ for the residual lensing B-mode $B'$, but for all the examples in this
section, we find that $C_\ell^{B'B'}$ is approximately constant on large scales, so that the residual B-mode can be treated as a source of white
noise in the forecast for $\sigma(T/S)$.}
While the precise values of $\sigma(T/S)$ achievable with and without delensing are difficult to forecast due to considerations such as loss of modes at low $\ell$
due to EB mixing
from survey boundaries\footnote{The most critical issue when forecasting $\sigma(T/S)$ is whether the gravity wave B-mode can be constrainted through the
reionization bump at $\ell\approx 8$, or whether only the recombination bump at $\ell\approx 60$ is measurable in the presence of foregrounds and sky cuts.  At the level of
a naive mode-counting forecast, the reionization bump has $\approx 10$ times the signal-to-noise of the recombination bump when 
constraining $(T/S)$.}  \cite{Lewis:2001hp,Bunn:2002df,Amarie:2005in,Smith:2006vq},
the ratio is simply given by:
\be
\frac{\sigma(T/S)_{\rm no\ delensing}}{\sigma(T/S)_{\rm with\ delensing}} = 
\frac{(\sigma^B_{\rm lensed})^2 + (\sigma^B_{\rm inst})^2}{(\sigma^B_{\rm delensed})^2 + (\sigma^B_{\rm inst})^2}  \label{eq:delensing_ratio}
\ee
For this reason, rather than presenting forecasts for $\sigma(T/S)$, we will forecast the ratio in Eq.~(\ref{eq:delensing_ratio}).
This ratio isolates the improvement in $\sigma(T/S)$ due to delensing alone, independent of the large-scale survey geometry and mode coverage.

\subsection{Delensing using small-scale polarization}
\label{ssec:polarization_delensing}

The first approach to delensing that we will consider is to construct the template $\hphi_{\ell m}$ for the lens potential ``internally'' from CMB polarization,
by applying a lens reconstruction estimator to the small-scale E and B-modes.
In polarization, the quadratic estimator $\hphi_{\ell m}$ that has been discussed previously (Eq.~(\ref{eq:hphi_def})) can be significantly improved for low noise levels
using an iterative, likelihood-based approach \cite{Hirata:2002jy,Hirata:2003ka}.
On an intuitive level, the improvement arises because lensed B-mode power acts as a source of noise for the quadratic estimator, but the estimated lens potential
can be used to ``delens'' and reduce the level of the lensing B-mode for a subsequent evaluation of the quadratic estimator, leading to an iterative estimator.
One qualitative difference between the two estimators is that if we consider the mathematical limit of zero instrumental noise (neglecting real-world issues such
as foregrounds and systematics), the quadratic estimator will have nonzero reconstruction noise, but the iterative estimator can reconstruct the lens potential $\phi$
and delens the B-mode perfectly.  In this idealized zero-noise limit there is no fundamental limit to the level of $(T/S)$ which can be detected \cite{Seljak:2003pn},
unlike the case of the quadratic estimator \cite{Knox:2002pe,Kesden:2002ku}.
Our forecasting methodology for delensing includes the improvements from using the iterative estimator, as described in App.~\ref{app:lens_reconstruction}.

\begin{figure}
\begin{center}
\includegraphics[width=4in]{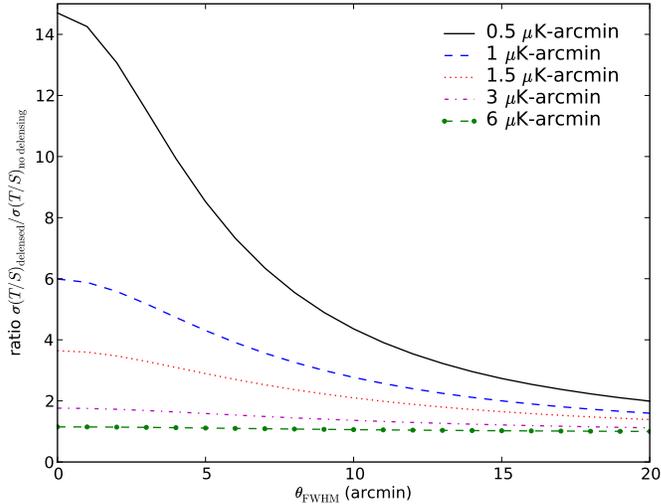}
\end{center}
\caption{Ratio of $\sigma(T/S)$ with and without polarization delensing, forecasted using Eq.~(\ref{eq:delensing_ratio}) for
varying instrumental noise level and beam.}
\label{fig:delensing_polarization}
\end{figure}

In Fig.~\ref{fig:delensing_polarization}, we show forecasts for the improvement in $\sigma(T/S)$ due to delensing (i.e. the ratio in Eq.~(\ref{eq:delensing_ratio}))
from an ``internal'' lens reconstruction using small-scale CMB polarization, for varying noise level and beam and taking $\ellmax=4000$ throughout.
Let us note some qualitative features of this figure.
For large beam size, delensing is not very effective since it depends on being able to reconstruct the lens potential indirectly through its effect on the small-scale
modes of the CMB.
The effective noise level for low-$\ell$ B-modes will simply be the sum of lensing ($\sigma^B_{\rm lensed} \approx 5$ $\mu$K-arcmin) and instrumental contributions.
As the beam size decreases, delensing can improve the lensing contribution, but the instrumental contribution is unchanged.
In the limit of a very small beam, the delensed noise level $\sigma^B_{\rm delensed}$ will reach an intermediate value which is less than $\sigma^B_{\rm lensed}$ but somewhat
larger than the instrumental noise $\sigma^B_{\rm inst}$.
For example, with 1 $\mu$K-arcmin instrumental noise and a 2' beam, we find $\sigma^B_{\rm delensed}=1.7$ $\mu$K-arcmin, resulting in a factor $\approx 7$ improvement in $\sigma(T/S)$
relative to the no-delensing case, as shown in Fig.~\ref{fig:delensing_polarization}.

Therefore, for B-mode experiments which are lensing-limited ($\sigma^B_{\rm inst} \lesssim 5$ $\mu$K-arcmin), the large and small scales are intimately linked: measuring
the gravity wave B-mode on large scales ultimately depends on reconstructing the lens potential via the lensing B-mode on large scales.
The effective noise level for constraining $(T/S)$ has a nontrival dependence on the beam size as shown in Fig.~\ref{fig:delensing_polarization}; if the noise level and beam size
are small, then large improvements in $\sigma(T/S)$ are possible.
In practice, since beam size is a primary driver of cost and complexity (particularly for a satellite mission), the improvement in $\sigma(T/S)$ which
we have forecasted in Fig.~\ref{fig:delensing_polarization} will be one factor to be weighed against others when designing an experiment.

\subsection{Delensing using small-scale temperature}
\label{ssec:temperature_delensing}

In this subsection and the next, we consider situations in which lensing-limited ($\lesssim 5$ $\mu$K-arcmin) CMB polarization measurements have been made on large angular scales, 
using an instrumental beam which is too large to observe the small-scale lensing B-mode needed for ``internal'' delensing.
Is it possible to delens the large-scale B-mode using ``external'' measurements from other datasets?
In order to perform delensing, we must have:
\begin{enumerate}
\item A (noisy) template $\hphi_{\ell m}$ for the CMB lens potential
\item A (noisy) measurement $E_{\ell m}$ of the CMB E-mode on intermediate ($\ell\lesssim 2000$) angular scales (e.g. from the Planck satellite \cite{planck}).
\end{enumerate}
The improvement in $\sigma(T/S)$ which can be achieved using delensing will depend on the noise levels in both $\hphi_{\ell m}$ and $E_{\ell m}$.

One possible way to get the template $\hphi_{\ell m}$
would be to apply the quadratic estimator to small-scale CMB {\em temperature} measurements from another experiment with high angular resolution.
Experiments are already underway (e.g. ACT \cite{Kosowsky:2006na} or SPT \cite{Ruhl:2004kv}) with sufficient sensitivity to measure the CMB temperature with high signal-to-noise,
far into the damping tail of the temperature power spectrum $(\ell > 2000)$.
For such experiments, the limiting factor in lens reconstruction is likely to be the presence of non-Gaussian secondary anisotropies (which become increasingly important as $\ell$
increases), rather than instrumental sensitivity or resolution.  However, at the time of this writing it is unclear what range of scales will be ``sufficiently Gaussian'' to use for lens
reconstruction in a real experiment \cite{Amblard:2004ih}.

\begin{figure}
\begin{center}
\includegraphics[width=4in]{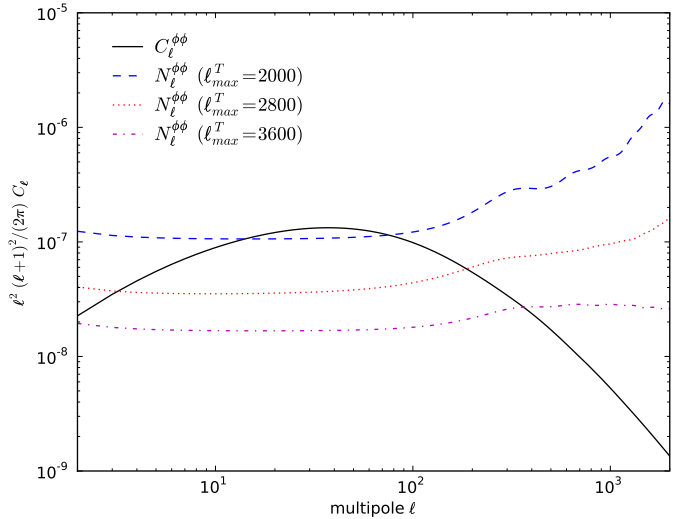}
\end{center}
\caption{Reconstruction noise power spectra $N_\ell^{\phi\phi}$ from CMB temperature alone, assuming cosmic variance limited measurements for varying values of $\ellmax^T$.}
\label{fig:temperature_nlphi}
\end{figure}

We will model this unclear situation in an approximate way by introducing a cutoff multipole $\ellmax^T$, and assuming that temperature multipoles $\ell\le\ellmax^T$ can be
used for lens reconstruction with the full statistical power of a Gaussian field (i.e. without introducing extra systematic error from secondary anisotropies), but multipoles
$\ell > \ellmax^T$ are not useful for lens reconstruction.
In Fig.~\ref{fig:temperature_nlphi}, we show noise power spectra $N_\ell^{\phi\phi}$ obtained using lens reconstruction from CMB temperature, for varying $\ellmax^T$.
As $\ellmax^T$ increases, a high signal-to-noise reconstruction is obtained on large scales, but on angular scales which are smaller than the CMB acoustic peak scale
($\ell\gtrsim 200$), the reconstruction always has poor signal-to-noise.

\begin{figure}
\begin{center}
\includegraphics[width=4in]{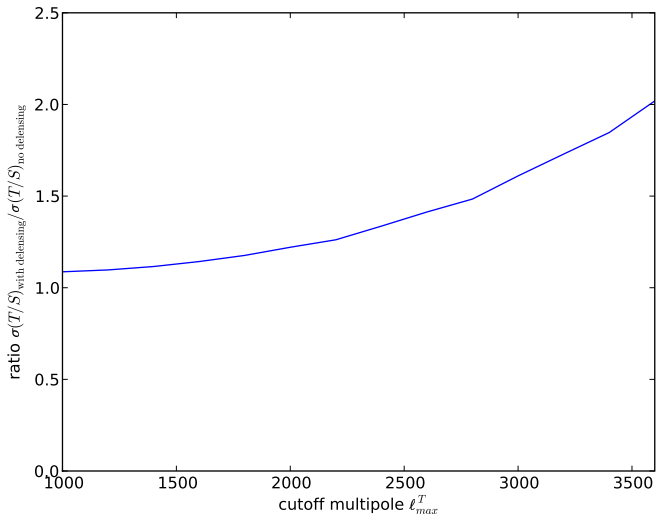}
\end{center}
\caption{Improvement in $\sigma(T/S)$ due to delensing (Eq.~(\ref{eq:delensing_ratio})) from CMB temperature alone, assuming cosmic variance limited measurements for varying values of $\ellmax$
(from \cite{delensing}).}
\label{fig:temperature_delensing}
\end{figure}

We would now like to forecast the improvement in $\sigma(T/S)$ due to delensing, i.e. the ratio in Eq.~(\ref{eq:delensing_ratio}).
In addition to the power spectrum $N_\ell^{\phi\phi}$ of the noise in the lensing template,
this ratio will depend on the noise $N_\ell^{EE}$ in the E-mode measurement on intermediate scales (i.e. item \#2 in the list above) and
the instrumental noise $\sigma^B_{\rm inst}$ on the large-scale B-mode.
However, an upper bound on the ratio can be obtained by neglecting these noise sources and assuming $N_\ell^{EE} = \sigma^B_{\rm inst} = 0$
(but keeping the nonzero $N_\ell^{\phi\phi}$ shown in Fig.~\ref{fig:temperature_nlphi}).
This upper bound is shown in Fig.~\ref{fig:temperature_delensing} for varying $\ellmax^T$ (taken from \cite{delensing}).
It is seen that, even for large $\ellmax^T$, the improvement is modest: a factor of two at $\ellmax^T=$3500.
In practice, this will be further degraded by the noise sources that have been neglected in obtaining this upper bound.
We interpret this as a negative result: it is not possible to delens CMB polarization using lens reconstruction from CMB temperature alone, because only the large-scale
modes in $\phi$ can be reconstructed with high signal-to-noise (Fig.~\ref{fig:temperature_nlphi}).

\subsection{Delensing using large-scale structure}
\label{ssec:lss_delensing}

We next consider another case of ``external'' delensing: using large-scale structure between the observer and recombination to obtain an external template $\hphi$.
One could imagine using different flavors of large-scale structure data (for example, cosmic shear \cite{Marian:2007sr} or 21-cm temperature 
\cite{Zahn:2005ap,Sigurdson:2005cp,Mandel:2005xh})
to construct $\hphi$, weighted to minimize the power spectrum of the residual field $(\hphi-\phi)$, where $\phi$ is the true CMB lens potential.

In the previous subsection, we obtained an upper bound on the improvement in $\sigma(T/S)$ that could be obtained using temperature multipoles $\ell\le\ellmax^T$,
by making some idealizing assumptions: we neglected noise in the in the E-mode on intermediate scales, and in the B-mode on large angular scales.
Since our final result showed only a modest improvement (Fig.~(\ref{fig:temperature_delensing})) even under these assumptions, we could intepret it as a general ``no-go'' theorem: lens
reconstruction from CMB temperature is of very limited utility in delensing the large-scale B-mode.
In this subsection, we will construct an analogous upper bound for the improvement in $\sigma(T/S)$ that can be obtained using measurements of large-scale structure
from redshifts $z\le\zmax$.

\begin{figure}
\begin{center}
\includegraphics[width=4in]{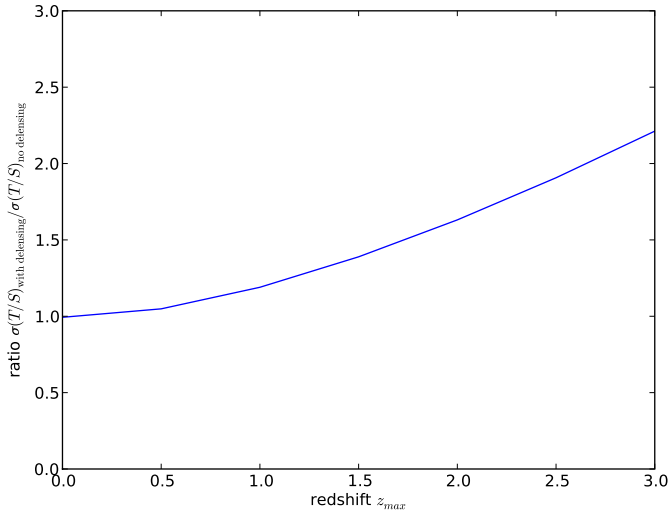}
\end{center}
\caption{Improvement in $\sigma(T/S)$ assuming perfect delensing of all large-scale structure from redshifts $\le \zmax$.}
\label{fig:lss_delensing}
\end{figure}

If we write the CMB lens potential $\phi$ as a line-of-sight integral (Eq.~(\ref{eq:line_of_sight})) with contributions from different redshifts, then by causality alone, 
contributions from redshifts $>\zmax$ cannot be reconstructed using large-scale structure, and must therefore be treated as ``noise'' power in the reconstruction.
(The reconstruction noise is defined to be the difference between the true CMB lens potential $\phi$ and the template $\hphi$ constructed from large-scale structure.)
More formally, a lower bound on the noise power spectrum $N_l^{\phi\phi}$ can be obtained by simply cutting off the redshift integral for $C_\ell^{\phi\phi}$ at redshift $\zmax$:
\be
N_\ell^{\phi\phi} \ge \frac{8\pi^2}{\ell^3} \int_{z_0}^\infty \frac{dz}{H(z)} D(z) P_\Psi(z,k=\ell/D(z)) \left( \frac{D(z_{\rm rec})-D(z)}{D(z_{\rm rec})D(z)} \right)^2  \label{eq:lss_lower_bound}
\ee
We now forecast the improvement in $\sigma(T/S)$ from delensing, making idealizing assumptions: we assume that the reconstruction noise is equal to the lower bound in
Eq.~(\ref{eq:lss_lower_bound}), and that $N_\ell^{EE}=\sigma^B_{\rm inst}=0$ as in the previous subsection.
The result is shown in Fig.~\ref{fig:lss_delensing} (taken from \cite{delensing}).
Since we find only a modest improvement (a factor of 2.2 for $\zmax=$3) even with our idealizations, we interpret this as a general negative result:
it is not possible to delens CMB polarization using large-scale structure.

This conclusion assumes that only large-scale structure from redshifts $\lesssim 3$ is available with sufficient statistical power to construct the template $\hphi_{\ell m}$.
One possible exception to this assumption may be a futuristic 21-cm experiment such as SKA or FFTT \cite{Tegmark:2008au}.
This possibility is studied in \cite{Sigurdson:2005cp} but is unlikely to be available for the next few generations of CMB polarization experiments.

\newpage
\section{Polarized foregrounds}
\label{sec:foregrounds}

Lens reconstruction relies on the non-Gaussian nature of the lensed CMB: each mode $\hphi_{\ell m}$ of the lens potential induces
a small deviation from Gaussian statistics, and this permits the potential to be reconstructed.
Because astrophysical foregrounds are non-Gaussian they are particularly worrisome for lens reconstruction.
For example, it is not clear {\em a priori} how to relate the ``strength'' of a foreground contaminant at the power spectrum level
to the bias that it produces in the reconstructed potential.
Detailed arguments which will be presented in \cite{HHRS} will show that polarized extragalactic point sources are expected to be the largest contaminant
of the reconstructed potential.
For this reason we will not analyze, e.g. polarized synchrotron or dust emission in this section, but we will do a detailed analysis of the contamination from
polarized point sources.
The extragalactic point sources can be modeled sufficiently well that a reliable estimate of the lens reconstruction bias can be made,
at least at a rough level.
This section is an abridged version of \cite{HHRS}, where more details will be given.

\subsection{Polarized point sources: forecasting machinery}

Naively, the contribution of point sources to reconstruction of the lensing power $\hC_{\ell}^{\phi\phi}$ is a four-point function in the locations and polarization angles of the sources. 
However, this picture simplifies if the polarization angles of distinct sources are assumed to be uncorrelated: the only terms which generate a nonzero expectation value are 1-source and 2-source terms.
In the absence of any observational evidence to the contrary, we will make this assumption.
It can then be shown \cite{HHRS} that the contribution of point sources to lensing reconstruction is contained in two effective power spectra
$C_\ell^{pp}, C_\ell^{p\phi}$ which are defined by:
\ba
\left\langle \sum_{ij} \left( 1 - \frac 12 \delta_{ij} \right) S_i^2 S_j^2 Y_{\ell m}(\n_i)^* Y_{\ell' m'}(\n_j) \right\rangle
&=&
C_\ell^{pp} \delta_{\ell\ell'} \delta_{mm'} \label{eq:ps_auto_def} \\
\left\langle \sum_i S_i^2 Y_{\ell m}(\n_i) \phi_{\ell'm'} \right\rangle 
&=& 
C_\ell^{p\phi} \delta_{\ell\ell'} \delta_{mm'}  \label{eq:ps_cross_def}
\ea
where $\langle\cdot\rangle$ denotes an average over realizations of the point source model, summation indices $i,j$ run over point sources in a given
realization, and $(\n_i,\ S_i)$ denote the location and {\em polarized} flux of source $i$.

The $C_{\ell}^{p\phi}$ power spectrum is due to the correlation of point sources with the lensing potential $\phi$. Extragalactic point sources are biased tracers of the large-scale matter distribution, and thus correlated with $\phi(\n)$ through the line-of-sight lensing integral of Eq. \eqref{eq:line_of_sight}. This leads to a bias in the reconstructed power spectrum given by
\be
\Delta \hC_{\ell}^{\phi \phi} = 
C_{\ell}^{\phi \phips}
=
\left( \frac{i}{2} \right)
\frac{C^{p\phi}_\ell}{2\ell+1}
\sum_{\ell_1\ell_2} \Lambda^{\phi*}_{\ell_1\ell_2\ell} \threej{\ell_1}{\ell_2}{\ell}{2}{-2}{0}
\ee
where we have defined:
\be
\Lambda^{\phi}_{\ell_1\ell_2\ell} = \sqrt{\frac{(2\ell_1+1)(2\ell_2+1)(2\ell+1)}{4\pi}} 
  \frac{(N_\ell)(\Gamma^{EB}_{\ell_1\ell_2\ell})}{(C_{\ell_1}^{EE}+N_{\ell_1}^{EE})(C_{\ell_2}^{BB}+N_{\ell_2}^{BB})}  \label{eq:lambdaphidef}
\ee

The $C_{\ell}^{pp}$ power spectrum, on the other hand, encapsulates the bias due to auto-correlations among the point sources. The relationship between the $C_\ell^{pp}$ and lensing bias is quite involved \cite{HHRS}. For the purposes of this report, however, we will be neglecting the auto-clustering of point sources and may set $\langle S_i^2 S_j^2 \rangle = \langle S_i^2 \rangle \langle S_j^2 \rangle$. In this case, $C_{\ell}^{pp}$ has the form
\be
C_\ell^{pp} = I_1 + I_2\cdot\delta_{\ell 0}.
\ee
We will refer to the biases originating from $I_1$ and $I_2$ respectively as the 1-pt and 2-pt Poisson terms. This form of $C_{\ell}^{pp}$ greatly simplifies the the calculation of the bias due to $C_{\ell}^{pp}$, and we find an additive effect on lens reconstruction given by
\ba
C_\ell^{\phips \phips} &=& \frac{I_1}{4} \sum_{\ell_1\ell_2\ell_1'\ell_2'} \frac{\Lambda_{\ell_1\ell_2\ell}^{\phi*} \Lambda_{\ell_1'\ell_2'\ell}^{\phi}}{(2\ell+1)^2}
                 \left[ 2 \threej{\ell_1}{\ell_2}{\ell}{2}{-2}{0} \threej{\ell_1'}{\ell_2'}{\ell}{2}{-2}{0} + \threej{\ell_1}{\ell_2}{\ell}{2}{2}{-4} \threej{\ell_1'}{\ell_2'}{\ell}{2}{2}{-4} \right] \nn \\
&+& \frac{I_2}{16\pi} \sum_{\ell_1\ell_2} \frac{(N_\ell)^2|\Gamma^\phi_{\ell_1\ell_2\ell}|^2}{(2\ell+1)(C_{\ell_1}^{EE}+N_{\ell_1}^{EE})^2(C_{\ell_2}^{BB}+N_{\ell_2}^{BB})^2}
\ea
Our picture of the point-source bias to lensing reconstruction thus consists of two terms:
\begin{itemize}
\item{A ``multiplicative'' bias $C_{\ell}^{\phi \phips}$ due to the cross-correlation of point sources with large-scale-structure.}
\item{An ``additive'' bias $C_{\ell}^{\phips \phips}$ due to the non-zero four-point function of the point sources. This in turn separates into 1-pt and 2-pt Poisson contributions.}
\end{itemize}
Our analysis of point source bias now has two remaining steps. First, we must estimate the power spectra $C_{\ell}^{pp}$ and $C_{\ell}^{p \phi}$ from current observational constraints on polarized sources; second, we forecast the point source bias given values of these power spectra. These steps are carried out in the following two subsections.

\subsection{Polarized point sources: modeling}
\label{sec:ps_modeling}

We will estimate the point source bias at a fiducial frequency $\nu=100$ GHz, under the following simplifying assumptions:
\begin{enumerate}
\item We will only consider radio point source contamination, assuming that the contribution from infrared sources is smaller or comparable
in order of magnitude at 100 GHz.
\item As previously discussed, we will ignore point source clustering, i.e. only consider the one-halo contribution to $C_\ell^{pp}$. 
\end{enumerate}
We expect the resulting estimate of the bias to be correct at the order-of-magnitude level, deferring a more detailed forecasting to future work \cite{HHRS}.

We will model the source counts of radio sources at 100 GHz using the fitting function from \cite{Waldram:2007eq}, based on extrapolating
multifrequency observations from 15--43 GHz \cite{Bolton:2004tw}:
\be
\frac{dN}{dS} = \frac{N_0}{S^\beta}  \label{eq:dnds_waldram}
\ee
where $\beta=2.15$ and $N_0=12$ Jy$^{1.15}$ sr$^{-1}$.
Fitting functions based on independent datasets have been also proposed in \cite{DeZotti:2004mn,Sadler:2007vt}
and agree within a factor $\approx 2$.

We will model the polarization fraction by assuming that the $Q,U$ components of the polarization are Gaussian distributed
and that the polarization angle is uniform distributed.  Under these assumptions, the PDF for the polarization is:
\be
\frac{dQ\,dU}{\pi\gamma^2S_T^2} \exp\left( -\frac{Q^2+U^2}{\gamma^2S_T^2} \right)   \label{eq:qu_pdf}
\ee
where we take RMS polarization fraction $\gamma=0.1$.
This choice is somewhat conservative; bright sources at 20 GHz are typically 1--5\% polarized but there is some observational evidence for an
increasing polarization fraction with decreasing flux \cite{Sadler:2006gd}.

To calculate the flux integrals of Eqs. (\ref{eq:ps_auto_def}, \ref{eq:ps_cross_def}), we will take a simplified view of the point source removal process. We suppose that all sources above some limiting flux $S^{\rm max}_T$ have been identified and masked with $100\%$ completeness, and those below are untouched. We further assume that this masking is performed in \textit{temperature}, as it is here that our current understanding of CMB source extraction is best developed. Neglecting the use of polarization data to mask sources is an additional conservative choice in our analysis. Although the signal strength is weaker in polarization by a factor of $\gamma^2$, the confounding CMB signal is smaller as well, and improvements in detection may be made. For a temperature threshold $S^{\rm max}_T$, the effective point source power spectra are given by
\ba
C_\ell^{pp} &=& \frac{N_0\gamma^4}{5-\beta} (S_{T}^{\rm max})^{5-\beta} + \left[ \frac{N_0\gamma^2}{3-\beta} (S^{\rm max}_T)^{3-\beta} \right]^2 \delta_{\ell 0}  \label{eq:cpp_tmask}\\
C_\ell^{p\phi} &=& -2b\gamma^2 \int_0^{S_T^{\rm max}} dS \int dz \left( \frac{d^2n}{dS\,dz} \right) \left( \frac{D_*-D(z)}{D_*D(z)^3} \right) P_{\Psi\delta}(r,k=\ell/D(z)) \label{eq:cpphi_tmask}
\ea
where $b$ denotes the (assumed constant) bias of the radio point sources, and the remaining notation follows Eq.~(\ref{eq:clphiphi}).

To compute the cross spectrum $C_\ell^{p\phi}$, we need a model for $(d^2n/dS\,dz)$, the joint flux-redshift distribution of the point sources.
We use the fitting function proposed in \cite{dunlop_peacock} (``RLF-1'', flat-spectrum) and extrapolate to 100 GHz 
by adding a (weakly $z$-dependent) normalization so that the source counts
$(dn/dS)$ are consistent with the power law in Eq.~(\ref{eq:dnds_waldram}).

\subsection{Polarized point sources: forecasts}
\label{sec:ps_forecasting}

We may now proceed to evaluate the magnitude of point source contamination for EB lensing reconstruction.

\begin{figure}
\begin{center}
\includegraphics[width=4in]{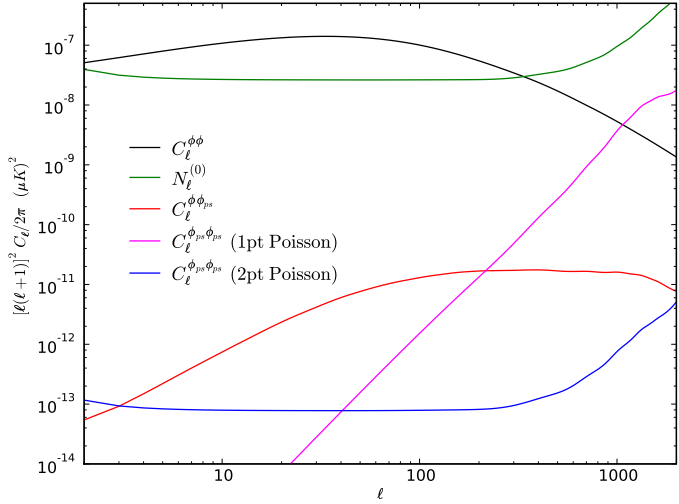}
\end{center}
\caption{Comparison of the signal power spectrum $C_\ell^{\phi\phi}$ and the bias terms in Eqs.~(\ref{eq:cpp_tmask}),~(\ref{eq:cpphi_tmask}), for an experiment with
$\Delta_P=4$ $\mu$K-arcmin, $\theta_{\rm FWHM}=7$ arcmin, $S_T^{\rm max}=200$ mJy.}
\label{fig:bias_sample}
\end{figure}

In Fig. \ref{fig:bias_sample}, we show the predicted bias for a $(7'\  {\rm FWHM}, \DeltaP=4\mu{\rm K-arcmin})$ experiment, with $S_{T}^{max}=200{\rm mJy}$. This is the $100\%$ completeness limit expected for the Planck 100GHz channel \cite{LopezCaniego:2006my}, and should therefore be readily achievable. For illustration, we plot the biases due to the components of Eqs.~(\ref{eq:cpp_tmask}),~(\ref{eq:cpphi_tmask}). Their behaviour here is characteristic of all of our forecasts. The 1-pt poisson term has a contribution which increases rapidly with $\ell$, and typically dominates by $\ell=1000$. The 2-pt poisson term is always subdominant, except at extremely low-$\ell$. The cross term $C_{\ell}^{p\phi}$ gives the largest contribution on intermediate scales. Note that these terms have different scalings with $S^{\rm max}$. The two poisson-type biases are due chiefly to point sources immediately below the removal threshold, whereas the $C_{\ell}^{p\phi}$ term receives contributions from all of the unresolved sources, and thus scales more slowly with $S^{\rm max}$.

To consider the effect of point source contamination more thoroughly we will need an estimate of the $S^{\rm max}$ which is achievable for a given experiment. The subject of point source extraction is an active one, with many techniques in active development \cite{BelenBarreiro:2005dv}. For summary purposes, however, we will take the following simplified model of this process. Suppose that we clean for point sources internally, by identifying all of the peaks in the CMB which are greater than $5\sigma$, relative to the total variance of the map. In temperature, the $\Delta\chi^{2}$ due to a single point source with a flux of $S_{T}$ is given by
\be
\Delta\chi^2 = \frac{S_T^2}{4\pi} \sum_\ell \frac{2\ell+1}{C_\ell^{TT}+N_\ell^{TT}}.
\ee
Solving for $\Delta\chi^2=25$ then gives
\be
S_T^{\rm max} = \left( \frac{1}{100\pi} \sum_\ell \frac{2\ell+1}{C_\ell^{TT}+N_\ell^{TT}} \right)^{-1/2}  \label{eq:temperature_pessimistic}.
\ee
This simple model gives values for the residual point source flux in reasonable agreement with those determined in more complete analyses \cite{LopezCaniego:2006my}. In line with our other assumptions, it is also a somewhat pessimistic estimate. It neglects, for example, our ability to increase the contrast of point sources by differencing CMB maps at multiple frequencies.

We also consider confusion as a lower limit on our ability to mask sources. Taking the differential number counts of Eq. \eqref{eq:dnds_waldram}, the number of point sources above a cutoff $S^{\rm max}$ is given by
\be
N(S > S_T^{\rm max}) = \frac{N_0 (S_T^{\rm max})^{1-\beta}}{\beta - 1}
\ee
To avoid confusion due to overlapping sources, we must ensure that the typical spacing between sources is $M$ beamwidths, for some reasonable value of $M$. We therefore require that
\be
S_{T}^{\rm max} \ge S_{\rm conf}^{\rm max}(M) = \left( \frac{N_0}{\beta - 1} \left[M \cdot \theta_{\rm FWHM}\right]^2 \right)^{\frac{1}{\beta-1}}
\label{eq:smaxt_confusion}
\ee
In what follows, we will take $M>10$. Satisfying Eq.~(\ref{eq:smaxt_confusion}) then ensures that $< {\cal O}(5\%)$ of the sky will be excised for point source removal. This has the benefit of limiting the issues with E/B mixing due to the masking process which we have otherwise neglected.

We would also like to establish a connection between biases at the power spectrum level which we calculate here and parameter constraints. For this, we use $m_{\nu}$ as a canary parameter, and plot the quantity
\be
\Delta_{m_{\nu}}(C_{\ell}^{\phi\phi}) = \left|\frac{\partial C_{\ell}^{\phi\phi}}{\partial m_{\nu}}\right| \cdot \sigma(m_{\nu})
\ee
If the bias to the power spectrum is well below this level, we expect that the effect on parameters will be negligible.

\begin{figure}
\begin{center}
\includegraphics[width=4.5in]{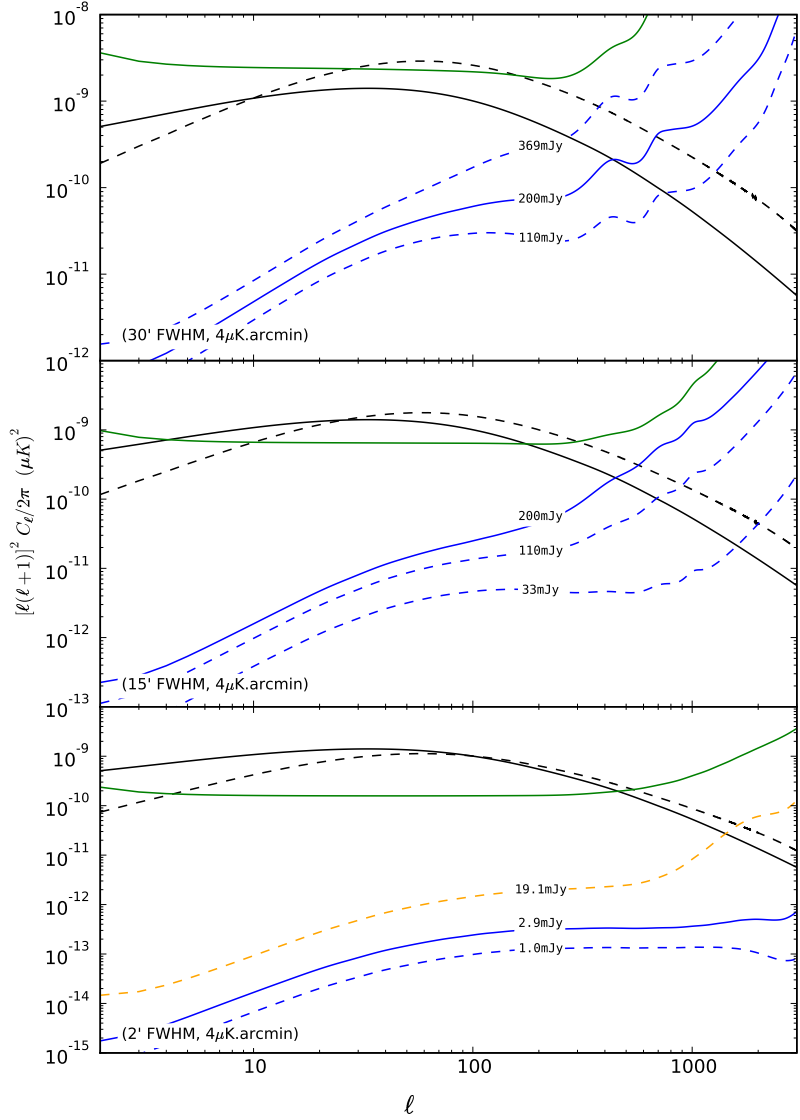}
\end{center}
\caption{Lensing reconstruction biases. Black/Green solid curves are $C_{\ell}^{\phi\phi}/N_{\ell}^{(0)}$ \textit{divided by 100}. The black dashed curves are $\Delta_{m_{\nu}}(C_{\ell}^{\phi\phi})$ (Eq. \ref{eq:smaxt_confusion}). The curves labelled in mJy are the total biases for the corresponding value of $S^{\rm max}_T$. Solid blue curves are for a `fiducial' $S^{\rm max}_T$-- the smaller of $200{\rm mJy}$ and the value determined from Eq. \eqref{eq:temperature_pessimistic}. The upper and lower dashed blue curves correspond to $S^{\rm max}_{\rm conf.}$ for $M=20$ and $M=10$ (only $M=10$ is shown for the $2'$ experiment for clarity, as the $M=20$ curve overlaps with the fiducial curve). The $S^{\rm max}_T=19.1{\rm mJy}$ curve for the $2'$ experiment is the value of Eq. \eqref{eq:temperature_pessimistic} assuming that only multipoles $\ell < 3000$ are used for source cleaning.}
\label{fig:30_15_2_biases.eps}
\end{figure}

In Fig. \ref{fig:30_15_2_biases.eps} we have plotted expected bias levels for $30'$, $15'$, and $2'\ {\rm FWHM}$ experimental configurations.
For all three beam sizes, we find that for reasonable values of $S^{\rm max}_{T}$, the point source bias generally small. Where it grows large enough to potentially bias parameter determinations, the (S/N) of the lens reconstruction is always $<0.1$, and so we expect that the potentially contaminated high-$\ell$ reconstruction may be disposed of without significant loss of information.

There are two caveats with these findings, however:
\begin{itemize}
\item{For a blunt ($30'\ {\rm FWHM}$) beam experiment, removal of point sources to the expected level may be hampered by the large beam size.}
\item{For a sharp ($2'\ {\rm FWHM}$) beam experiment, our findings are particularly dependent on the source count relation of Eq. \eqref{eq:dnds_waldram} holding true to at fluxes of $<20 {\rm mJy}$, where complete measurements have not yet been made.}
\end{itemize}
Exempting these two possible issues, to the best of our current modeling ability it appears that radio point sources will not constitute a significant difficulty for a future polarization based lensing reconstruction.

\newpage
\section{Beam systematics}
\label{sec:systematics}

To calculate the effect of beam systematics we invoke the Fisher 
information-matrix formalism.
Our objective is to determine the susceptibility of certain 
cosmological parameters to beam systematics.
We represent the extra noise 
due to beam systematics by analytic approximations \cite{Shimon:2007au}
and include lensing extraction in the parameter inference process, 
following \cite{Kaplinghat:2003bh,Lesgourgues:2005yv}
for neutrino mass (and other cosmological parameters) inference from CMB data.

Our main concern is the effect on the tensor-to-scalar ratio $r$ 
and the total neutrino mass $M_{\nu}$ (assuming three degenerate species).
The lensing-induced B-mode signal is sensitive to neutrino masses and therefore 
a large enough beam systematic which leaks temperature or E-mode polarization 
to B-mode polarization can bias the inferred neutrino mass.

\subsection{General}

Beam systematics due to optical imperfections depend on both the underlying sky, 
the properties of the polarimeter and on the scanning strategy. 
An instructive example is the effect of differential pointing. This effect depends on 
the temperature gradient to first order. CMB temperature gradients at the $1^{\circ}$, $30'$, 
$10'$, $5'$ and $1'$ scales are $\approx$ 1.4, 1.5, 3.5, 2.5 and 0.2 $\mu K/{\rm arcmin}$, respectively.
Therefore, any temperature difference measured with a dual-beam experiment with
a $\approx 1'$ pointing error and non-ideal scanning strategy which is dominated by its dipole and 
octupole moments \cite{Shimon:2007au} will result in a $\approx 1\mu$K systematic 
polarization which has the potential to contaminate the B-mode signal. 

Similarly, the systematic induced by differential ellipticity results from the variation 
of the underlying temperature anisotropy along the two polarization-sensitive directions 
which, in general, differ in scale depending on the mean beamwidth, degree of ellipticity 
and the tilt of the polarization-sensitive direction with respect to the ellipse's 
principal axes. For example, the temperature difference measured along the major and minor axes 
of a $1^{\circ}$ beam with a 2\% ellipticity scales as the second gradient of the 
underlying temperature which on this scale is $\approx 0.2\mu K/{\rm arcmin^{2}}$ 
and the associated induced polarization is therefore expected to be on the $\approx\mu K$ 
level.
The spurious signals due to pointing error, 
differential beamwidth and beam ellipticity all peak at angular scales comparable to the beam size 
(since they are associated with features in the temperature anisotropy on sub-beam scales on the 
one hand but suffer beam dilution on yet smaller scales). If the beam 
size is $\approx 1^{\circ}$ the beam systematics mainly affect the deduced tensor-to-scalar ratio, $r$.
If the polarimeter's beamwidth is a few arcminutes the associated systematics will impact the measured 
neutrino mass $m_{\nu}$, spatial curvature $\Omega_{k}$, running of the scalar spectral index 
$\alpha$ and the dark energy equation of state $w$ (which strongly affects the lensing-induced B-mode signal). 
It can certainly be the case that other cosmological parameters will be affected as well.

Two other spurious polarization signals we explore are due to differential gain and differential rotation; 
these effects are associated with different beam `normalizations' and orientation, respectively, and 
are independent of the coupling between beam substructure and the underlying temperature perturbations.
In particular, they have the same scale dependence as the primordial 
temperature anisotropy and polarization power spectra, respectively, and their peak impact 
will be on scales associated with the CMB's temperature anisotropy ($\approx 1^{\circ}$) 
and polarization ($\approx 10'$).

\subsection{The effect of systematics on lens reconstruction}

Gravitational lensing of the CMB is both a nuisance and a valuable cosmological tool 
(e.g. \cite{Zaldarriaga:1998ar}). 
It certainly has the potential to complicate CMB data analysis due to the non-gaussianity 
it induces. However, it is also a unique probe of the growth of structure in the 
linear, and mildly non-linear, regimes (redshift of a few). 
In \cite{Kaplinghat:2003bh,Lesgourgues:2005yv} as well as elsewhere
it was shown that with a {\it nearly ideal 
CMB experiment}, neutrino mass limits can be improved by a factor of approximately four
by including lensing extraction in the data analysis using CMB data alone.

This lensing extraction process is not perfect; residual noise will 
affect any experiment, even an ideal one.
This noise will, in principle, propagate to the inferred cosmological parameters if 
the latter significantly depend on lensing extraction, e.g. neutrino mass, $\alpha$ and $w$.
It is important to illustrate first the effect of beam systematics on lensing reconstruction.
By optimally filtering the temperature and polarization, the lens potential can be recovered
using quadratic estimators \cite{Carroll:1998zi}.
It was shown that for experiments with ten times higher sensitivity than 
Planck, the EB estimator yields the tightest limits on the lens potential. This 
conclusion assumes no beam systematics which might significantly contaminate the observed 
B-mode. 

We illustrate the effect of differential gain, beamwidth, beam rotation, 
ellipticity and differential pointing (see \cite{Shimon:2007au}) on the noise of lensing 
reconstruction with 
20 different combinations of noise levels and resolution for CMBPOL 
spanning the sensitivity and resolution ranges 1-6 ${\rm \mu K}$-arcmin 
and 3-30 arcmin, respectively.
These are perhaps the most pernicious systematics. Beam rotation  
induces cross-polarization which leaks the much larger 
E-mode to B-mode polarization and differential ellipticity leaks T to B. 
The modified noise in reconstructing the lens potential, $N_{l}^{dd}$, is 
consistently substituted into our Fisher-matrix analysis. 

\subsection{Error forecast}

Accounting for beam systematics in both Stokes parameters and lensing 
power spectra is straightforward. In addition, the detector noise associated with 
the main beam is accounted for, as is conventional, 
by adding an exponential noise term. Assuming gaussian white noise
\begin{eqnarray}
N_{l}=\frac{1}{\sum_{a}(N_{l}^{aa})^{-1}}
\end{eqnarray}
where $a$ runs over the experiment's frequency bands. The noise in channel {\it a} is 
(assuming a gaussian beam) 
\begin{eqnarray}
N_{l}^{aa}\equiv W_{T}^{-1}e^{l(l+1)\theta_{a}^{2}/8\ln(2)}
=(\theta_{a}\Delta_{a})^{2}e^{l(l+1)\theta_{a}^{2}/8\ln(2)},
\end{eqnarray}
where $\Delta_{a}$ is the noise per beam in $\mu$K,
$\theta_{a}$ is the beam width,
and we assume noise from different channels is uncorrelated.
The power spectrum then becomes
\begin{eqnarray}
C_{l}^{X}\rightarrow C_{l}^{X}+N_{l}^{X}
\end{eqnarray}
where $X$ is either the auto-correlations $TT$, $EE$ and $BB$ or the cross- 
correlations $TE$, $TB$ and $EB$ (the latter two power spectra vanish in the 
standard model but not in the presence of beam systematics and exotic parity-violating physics (e.g. \cite{Carroll:1998zi,Liu:2006uh,Xia:2007qs,Komatsu:2008hk})
or primordial magnetic fields (e.g. \cite{Kosowsky:1996yc}). 
For the cross-correlations, the $N_{l}^{X}$ vanish as 
there is no correlation between the instrumental noise of the 
temperature and polarization (in the absence of beam systematics).

\subsection{Results}

We consider the effect of both irreducible and reducible systematics. By `reducible' we refer to systematics which depend on the coupling of an imperfect scanning strategy to the beam mismatch parameters. 
These can, in principle, be removed or reduced during data analysis.
This includes the differential gain, differential beamwidth and first order pointing error beam systematics. By `irreducible' we refer to those systematics that depend only on the beam mismatch parameters (to leading order). For instance, the differential ellipticity and second order pointing error persist even if the scanning strategy is ideal. For reducible systematics the scanning strategy 
is a free parameter in our analysis (under the assumption it is non-ideal, yet uniform, over the map)
and we set limits on the product of the scanning strategy (encapsulated by the $f_{1}$ and $f_{2}$ parameters) and the differential gain, beamwidth and pointing, as will be described below.
The exact definitions of $f_{1}$ and $f_{2}$ are given in \cite{Shimon:2007au,Miller:2008zi} but here we give approximate expressions 
under the assumption that the scanning strategy does not contain significant hexadecapole moment.
\begin{eqnarray}
f_{1}&=&\frac{1}{2}|A|^{2}\nonumber\\
f_{2}&=&\frac{1}{2}(|B|^{2}+|C|^{2}) 
\end{eqnarray}
where
\begin{eqnarray}
A&\approx &\langle\exp (2i\alpha)\rangle\nonumber\\
B&\approx &\langle\exp (3i\alpha)\rangle\nonumber\\
C&\approx &\langle\exp (i\alpha)\rangle
\end{eqnarray}
and therefore while $f_{1}$ captures the quadrupole moment of the scanning strategy, $f_{2}$ is a 
measure of its dipole and octupole moments. To calculate the power spectra 
we assume the concordance cosmological model throughout; the baryon, cold dark mater, 
and neutrino physical energy densities in critical density units $\Omega_{b}h^{2}=0.021$, 
$\Omega_{c}h^{2}=0.111$, $\Omega_{\nu}h^{2}=0.006$. The latter is equivalent to a total neutrino mass $M_{\nu}=\sum_{i=1}^{3}m_{\nu,i}=$0.56eV, slightly lower than the current limit set by a joint analysis of the WMAP data and a variety of other cosmological probes (0.66eV, e.g. \cite{Spergel:2006hy}). We assume degenerate neutrino masses, i.e. all neutrinos have the same mass, 0.19 eV, for the purpose of illustration, and we do not attempt to address here the question of what tolerance levels are required to determine the neutrino hierarchy. As was shown in \cite{Lesgourgues:2005yv}, the prospects for determining the neutrino hierarchy from the CMB {\it alone}, even in the absence of systematics, are not very promising. 
This conclusion may change when other probes, e.g. Ly-$\alpha$ forest, are added to the analysis.
Dark energy makes up the rest of the energy required for closure density. 

We limit our analysis to the tensor-to-scalar ratio $r$, 
and total neutrino mass $M_{\nu}$. While $r$ is mainly constrained by the primordial B-mode signal that peaks on degree scales (and is therefore not expected to be overwhelmed by the beam systematics which peak at sub-beam scales), it is still susceptible to the tail of these systematics, extending all the way to degree scales, because of its expected small amplitude (less than $0.1\mu K$). The tensor-to-scalar ratio is also affected by differential gain and rotation which are simply rescalings of temperature anisotropy and E-mode polarization power spectra, respectively, and therefore do not necessarily peak at scales beyond the primordial signal. 

Ideally, the lensing signal, which peaks at $l\approx 1000$, provides a useful handle on the neutrino mass as well as other cosmological parameters which govern the evolution of the large scale structure and gravitational potentials. However, the inherent noise in the lensing reconstruction process \cite{Hu:2001kj} which depends, among other things, on the instrument specifications (detector noise and beamwidth), now depends on beam systematics as well. The systematics, however, depend on the cosmological parameters through temperature leakage to polarization, and as a result there is a complicated interplay between these signals and the information they provide on cosmological parameters. As our numerical calculations show, the effect on the inferred cosmological parameters stems from both the {\it direct effect} of the systematics on the parameters 
and the {\it indirect effect} on the noise in the lensing reconstruction, $N_{l}^{dd}$, in cases where the MV estimator is dominated by the EB correlations.

The Fisher information-matrix gives a first order approximation to the lower bounds on errors inferred for these parameters. 
We follow \cite{O'Dea:2006di} in quantifying the required tolerance on the differential gain, differential beamwidth, pointing, ellipticity and rotation. To estimate the effect of systematics and to set the systematics to a given tolerance limit one has to compare the systematics-free $1\sigma$ error in the i-th parameter to the error obtained in the presence of systematics. The latter has two components; the bias and the uncertainty (which depends on the curvature of the likelihood function, i.e. to what extent does the information matrix constrain the cosmological model in question). As in \cite{O'Dea:2006di} we define
\begin{eqnarray}
\delta &=&\frac{\Delta\lambda_{i}}{\sigma_{\lambda_{i}}}|_{\lambda_{i}^{0}}\nonumber\\
\beta &=&\frac{\Delta\sigma_{\lambda_{i}}}{\sigma_{\lambda_{i}}}|_{\lambda_{i}^{0}} 
\end{eqnarray}
where the superscript $0$ refers to values evaluated at the peak of the likelihood function, i.e. the values we assume for the underlying model, and $\Delta\lambda_{i}$ and $\Delta\sigma_{\lambda_{i}}$ are the bias and the change in the statistical error for a given experiment and for the parameters $\lambda_{i}$ induced by the beam systematics, respectively. As shown in \cite{O'Dea:2006di} these two parameters depend solely on the primordial, lensing and systematics power spectra. We require both $\delta$ and $\beta$ not to exceed 10\% of the uncertainty without systematics. 

\subsection{Expected Beam Uncertainties}

Before quoting and discussing the allowed levels of differntial ellipticity, gain and beamwidth it is instructive 
to estimate the uncertainty within which these beam parameters will be determined from a beam-calibration with 
a nearly black-body point-source such as Jupiter ($T_{p}\approx 200K$, $\theta_{p}\approx 0.5$ arcmin).
By Wiener filtering
a map of the observed source one expects to recover the source image with a signal-to-noise level 
\begin{eqnarray}
\left(\frac{S}{N}\right)^{2}
=\int\frac{|\tilde{S}({\bf l})|^{2}}{P({\bf l})}\frac{d^{2}{\bf l}}{(2\pi)^{2}}.
\end{eqnarray}
where $\tilde{S}({\bf l})$ is the Fourier transform of the point source and $P(l)$ is the instrumental noise, 
i.e. the $(S/N)^{2}$ is the ratio of the signal and noise power-spectra integrated over all accessible multipoles in the experiment in question. The higher S/N the smaller are the uncertainties in the recovered beam parameters.
The Fourier transform of the convolved point source reads
\begin{eqnarray}
\tilde{T}_{p}^{\rm obs}=(1+g)\tilde{T}_{p}
e^{-\frac{1}{2}l_{x}^{2}\sigma_{x}^{2}-\frac{1}{2}l_{y}^{2}\sigma_{y}^{2}-i{\bf l}\cdot\rho}
\end{eqnarray}
where we assume an elliptical gaussian beam with principal axes $\sigma_{x}$ and $\sigma_{y}$, gain $1+g$ and 
pointing ${\bf \rho}$. This results in 
\begin{eqnarray}
\left(\frac{S}{N}\right)^{2}=\frac{(1+g)^{2}2\ln(2)}
{\pi (1-e^{2})}\left(\frac{T_{p}}{\Delta_{b}}\right)^{2}\left(\frac{\theta_{p}}{\theta_{b}}\right)^{4}\eta^{2}.
\end{eqnarray}
Since the pointing merely adds a phase to the beam function it drops from the expression for S/N. Similarly, 
S/N is also independent on the beam rotation angle since temperature measurements are insensitive to $\varepsilon$. 
$\eta$ is an experiment-specific optical-efficiency parameter ($\le 1$). Therefore, the following procedure, which is based on S/N considerations, will be used to determine the uncertainty of $e$, $\mu$ and $g$ only. To determine these uncertainties we require that varying the beam parameters will result in signal changes smaller than the noise
\begin{eqnarray}
(S/N)^{2}\rightarrow (S/N)^{2}+1.  
\end{eqnarray}
This condition readily yields the uncertainty in beam parameters
\begin{eqnarray}
\Delta (e^{2})=\Delta (\mu^{2})=\Delta (g^{2})=[S/(N\eta)]^{-2}. 
\end{eqnarray}
For each `experiment' of the 20 considered here we plug in $\Delta_{b}$ and $\theta_{b}$ and obtain 
$\eta N/S$. Table 3 summarizes $\eta N/S$ in \%-units (as in Miller, Shimon \& Keating 2008). The color-coded threshold values reported in Tables 4, 6 \& 7 (differential gain, beamwidth and ellipticity, respectively) should be compared with those of Table 3. Values in green are those which meet the requirements from the uncertainties (reported in Table 3) by better than factor 20. 
Blue figures are those which meet the fundamental uncertainties specified in Table 3 in case the optical efficiency is $\eta=1$ but fail to do so if it is as low as $\eta=0.05$.  

\subsection{Conclusions}

The upper limits we obtained on the allowed range of beam mismatch parameters for given experiments and given arbitrarily-set tolerance levels on the parameter bias and uncertainty, constitute conservative limits in 
the treatment of systematics but on the other hand they neglect potential confusion sources with lensing in the worst case scenarios as we explain below.
It may be the case that a few of the systematics studied here may be fully or partially removed. This includes, in particular, the first order pointing error which couples to the dipole and octupole moments of non-ideal scanning strategies (see \cite{Shimon:2007au}). By removing this dipole during data analysis the effect due to the systematic first order pointing error (dipole) may drop dramatically. We made no attempt to remove or minimize these effects in this work. Our results highlight the need for scan mitigation techniques because the coupling of several beam systematics to non-ideal scanning strategies result in systematic errors. This potential solution may reduce systematics, which ultimately propagate to parameter estimation, and affect mainly the parameters to which the B-mode polarization is sensitive. A brute-force strategy to idealize the data could be to remove data points that contribute to higher-than-the-monopole moments in the scanning strategy.
This would effectively make the scanning strategy `ideal' and alleviate the effect of the {\it a priori} most pernicious beam systematics. 
This procedure `costs' only a minor increase in the instrumental noise (due to throwing out a fraction of the data) 
but may potentially reduce the most pernicious reducible beam systematic, i.e. the first order pointing error (`dipole' effect). 

Our results are summarized in Tables~4--8, where we list upper limits on the allowed beam systematics 
(differential gain, pointing, beamwidth, ellipticity and rotation, respectively) for various combinations of CMBPOL noise and resolution parameters 
(in units as in \cite{Miller:2008zi}, i.e. allowed differential gain, pointing, beamwidth, ellipticity and beam rotation are given in $\sqrt{\frac{f_{1}}{2\pi}}$\%, $\sqrt{\frac{f_{2}}{2\pi}}$arcsec, $\sqrt{\frac{f_{1}}{2\pi}}$\%, \% and degrees, respectively) based on the requirement that the bias induced in the tensor-to-scalar ratio $r$ (first value) and $M_{\nu}$ (second value) do not exceed the 10\% level. 
Before describing specific results for the various systematics and cosmological parameters 
$r$ and $M_{\nu}$ we comment that all threshold values found from our analysis for the differential gain, beamwidth and ellipticity are larger than the beam uncertainties obtained in section 5.5 (Table 3) and therefore the beam systematics meet the requirements.

As for the pointing and beam rotation; these are unconstrained by the beam calibration with unpolarized point-source as was illustrated in section 5.5. However, as explained above, the effect of pointing may be harnessed by removing non-ideal moments of the scanning strategy and addressing the effect of polarization-mixing will require an accurate measurement of polarization direction.
Tables~4 and~5
show the tolerance for the differential gain and pointing respectively, subject to the bias in r and $M_{\nu}$. 
Our discussion begins with $r$. As expected, when the sensitivity and resolution of the experiment increase - the bounds on the allowed systematics are more demanding. 
Tables~6 and~7
refer to the differential beamwidth and ellipticity. Here the dependence on sensitivity is as before but the allowed systematics actually increase as the angular resolution improves. The reason for this behavior is simple; both differential beamwidth and ellipticity scale as second gradients of temperature. As a result, they steeply rise as a function of multipole number and effectively peak on scales smaller than the beamsize. This implies, for example, that if we are interested mainly in the tensor-to-scalar ratio $r$ we should consider having our beams very narrow so as to push the systematic signal to small angular scales, beyond the inflationary peak at $\sim 2^{\circ}$. 
Table~8, which describes the allowed beam rotation based on the requirement on $r$ is consistent with the general picture we saw with the differential gain and pointing 
(Tables~4 and~5); increasing resolution and sensitivity implies stronger limits on the allowed systematics levels.

For $M_{\nu}$ forecasts the picture is more complicated mainly due to several competing effect and the fact that lensing extraction benefits most from few arcminute scales and few of beam systematics peak on sub-beam scales.
Higher sensitivity experiments (with 1 and 2${\rm \mu K-arcmin}$) exhibit interesting behavior: the most stringent limits come from 5'-20' resolution experiments. This is where the B-mode from lensing peaks and since these low-noise experiments are sensitive to the lensing signal even small systematics might potentially bias the inferred $M_{\nu}$.

Table~5 describes the allowed pointing levels. A low sensitivity experiment (6 ${\rm \mu K}$-arcmin) allows increasing levels of differential pointing 
as we increase the resolution.
However, as we increase the sensitivity the most stringent constraints come from experiments with $\sim 10'$ resolution as the lower noise level allows to `see' larger portions of the B-mode lensing signal and therefore the allowed systematics are relatively smaller. Now, for a given resolution; moderate-low resolution experiments (10'-30'), which do not target the peak of the B-mode lensing signal even with systematics-free experiment, allow increasing levels of pointing as we `turn on' the instrumental noise. When the resolution is relatively high (3'-5') we still obtain increase of allowed systematics with increasing noise but this increase is not monotonic, rather - there is a distinguishable `dip' at around 4${\rm \mu K}$-arcmin: a possible explanation is that there are two competing effects in action. The first is that increasing the instrumental noise naturally allows increasing systematics without significantly affecting the uncertainty on inferred parameters. On the other hand increasing the instrumental noise limits lensing extraction and therefore relatively increases the weight of the information contained in the primary signal. The latter also suffers from contamination of E- and B-modes by beam systematics. The interpretation of the ellipticity constraints (Table~7) is similar to that of Table~6.
 Finally, the `undulations' in the allowed beam rotation, as shown in Table~8, merely reflect the relative ratios of primordial E- and lensing-induced B-mode at different multipoles accessible at the various resolutions.

As seen from the above, the allowed beam asymmetry parameters are non-trivial functions of both sensitivity and angular resolution as well as the cosmological parameter in question; $r$ or $M_{\nu}$. They strongly depend on the tolerance criterion (whether it is $r$ that depends on the primordial B-mode signal at angular degree scales or $M_{\nu}$ which is extracted largely from the higher multipole regime - lensing extraction from the few-arcmin B-mode signal). A possibly important factor in this context is the typical angular scales of these various types of systematics (i.e their $l$-dependence).

Throughout this study we invoked the standard quadratic estimators for lensing reconstruction by Hu \& Okamoto - this is allowed since we assume the scanning strategy is uniform across the sky (and as a result it induces no new typical scale and hence no new non-gaussianity). In practice however, this need not necessarily be the case; there are new non-gaussianities induced by the coupling of scanning strategy to the underlying sky and beam asymmetry. Consequently, the quadratic estimators will be biased and a more thorough, Monte-Carlo-Based study, needs to be carried out in order to fully address this issue. The effect of differential pointing, in particular, may mimic lensing by `shifting' features in the polarization maps and since its leading order contribution depends on the coupling of temperature anisotropy, beam pointing and scanning strategy, the later will cause a mode-mode coupling if it is non-uniform, inducing non-gaussianities. However, this very property of coupling to scanning strategy can be used, in principle, to remove it, at least partially.
In addition, the data may be uniformized to some degree by throwing out data points which contribute to the non-ideal scanning strategy (see section 5.4).

\begin{table*}
\begin{center}
\begin{tabular}[c]{|c|c|c|c|}
\hline
~{\rm depends on} ~&~effect~&~parameter~&~definition~\\
~{\rm beam substructure}~&~~&~~&~~\\
\hline
~{\rm No}~&~gain~& $g$ & $g_{1}-g_{2}$ \\
\hline
~{\rm Yes}~&~monopole~& $\mu$ & $\frac{\sigma_{1}-\sigma_{2}}{\sigma_{1}+\sigma_{2}}$ \\
\hline
~{\rm Yes}~&~dipole~& $\rho$ & ${\bf \rho}_{1}-{\bf \rho}_{2}$ \\
\hline
~{\rm Yes}~&~quadrupole~& $e$ & $\frac{\sigma_{x}-\sigma_{y}}{\sigma_{x}+\sigma_{y}}$ \\
\hline
~{\rm No}~&~rotation~& $\varepsilon $ & $\frac{1}{2}(\varepsilon_{1}+\varepsilon_{2})$ \\
\hline
\end{tabular}
\caption{Definitions of the parameters associated with the systematic effects.
Subscripts 1 and 2 refer to the first and second polarized beams of the dual beam polarization assumed in this work.} 
\end{center}
\end{table*}

\begin{table*}
\begin{center}
\begin{tabular}{|c|c|c|c|c|c|}
\hline
~effect~&~parameter~&~$\Delta C_{l}^{TE}$~&~$\Delta C_{l}^{E}$~&~$\Delta C_{l}^{B}$~\\
\hline
~gain~& $g$ & 0 & $g^{2}f_{1}\star C_{l}^{T}$ & $g^{2}f_{1}\star C_{l}^{T}$\\
\hline
~monopole~& $\mu$ & 0 & $4\mu^{2}(l\sigma)^{4}C_{l}^{T}\star f_{1}$ & $4\mu^{2}(l\sigma)^{4} C_{l}^{T}\star f_{1}$\\
\hline
~pointing~& $\rho$ & $-c_{\theta}J_{1}^{2}(l\rho)C_{l}^{T}\star f_{3}$ & 
$J_{1}^{2}(l\rho)C_{l}^{T}\star f_{2}$ & $J_{1}^{2}(l\rho)C_{l}^{T}\star f_{2}$\\
\hline
~quadrupole~& e & $-I_{0}(z)I_{1}(z)c_{\psi}C_{l}^{T}$ & 
$I_{1}^{2}(z)c_{\psi}^{2}C_{l}^{T}$ & $I_{1}^{2}(z)s_{\psi}^{2}C_{l}^{T}$ \\
\hline
~rotation~& $\varepsilon$ & $0$ & $4\varepsilon^{2}C_{l}^{B}$ & $4\varepsilon^{2}C_{l}^{E}$\\
\hline
\end{tabular}
\caption{The scaling laws for the systematic effects to 
the power spectra $C_{l}^{T}$, $C_{l}^{TE}$, $C_{l}^{E}$ and $C_{l}^{B}$ 
assuming the underlying sky is not polarized (except 
for the {\it rotation} signal where we assume the E, and B-mode signals are present) 
and a general, not necessarily ideal or uniform, scanning strategy. The next order contribution 
($~10\%$ of the `pure' temperature leakage shown in the table) is contributed by $C_{l}^{TE}$. 
It can be easily calculated based on the general expressions in \cite{Shimon:2007au} where 
the definitions of $z$, $\rho$, $\varepsilon$, etc., are also found. 
For the pointing error we found that the `irreducible' contribution to B-mode contamination, arising from a second order effect, is extremely small and therefore only the first order terms (which vanish in ideal scanning strategy) are shown. The functions $f_{1}$ and $f_{2}$ are experiment-specific and encapsulate the information about the scanning strategy which couples to the beam mismatch parameters to generate spurious polarization. In general, the functions $f_{1}$ and $f_{2}$ are spatially-anisotropic but for simplicity, and to obtain a first-order approximation, we consider them  constants in general. In the case of ideal scanning strategy they identically vanish. The exact expressions are given in \cite{Shimon:2007au}.} 
\end{center}
\end{table*}

\begin{table*}
\begin{center}
\begin{tabular}[c]{|c|c|c|c|c|}
\hline
~~&~1~&~2~&~4~&~6~\\
~~&~[$\mu$K-arcmin]~&~[$\mu$K-arcmin]~&~[$\mu$K-arcmin]~&~[$\mu$K-arcmin]~\\
\hline
~3'~&~6.13e-5~&~1.23e-4~&~2.45e-4~&~3.68e-4~\\
\hline
~5'~&~1.03e-4~&~2.04e-4~&~4.09e-4~&~6.13e-4~\\
\hline
~10'~&~2.04e-4~&~4.09e-4~&~8.16e-4~&~1.23e-3~\\
\hline
~20'~&~4.09e-4~&~8.16e-4~&~1.64e-3~&~2.45e-3~\\
\hline
~30'~&~6.13e-4~&~1.23e-3~&~2.45e-3~&~3.68e-3~\\
\hline
\end{tabular}
\caption{Epected uncertainty in the beam parameters $e$, $g$ and $\mu$ ($=\eta N/S$) from beam calibration with point-like source Jupiter ($\theta_{p}\approx 0.5$ arcmin), all given in \% units, as a function of instrument sensitivity and beamwidth (as described in section 5.5). $\eta$ ($ < 1$) is the experiment-specific optical efficiency; this parameter encapsultes our {\it current} ignorance of the experiment optics.}
\end{center}
\end{table*}

\begin{table*}
\begin{center}
\begin{tabular}[c]{|c|c|c|c|c|}
\hline
~~&~1~&~2~&~4~&~6~\\
~~&~[$\mu$K-arcmin]~&~[$\mu$K-arcmin]~&~[$\mu$K-arcmin]~&~[$\mu$K-arcmin]~\\
\hline
~3'~&~\textcolor{green}{0.00155}~&~\textcolor{blue}{0.00245}~&~\textcolor{blue}{0.00384}~&~\textcolor{blue}{0.00520}~\\
~&~\textcolor{green}{0.029}~&~\textcolor{green}{0.054}~&~\textcolor{green}{0.100}~&~\textcolor{green}{0.150}~\\
\hline
~5'~&~\textcolor{blue}{0.00156}~&~\textcolor{blue}{0.00246}~&~\textcolor{blue}{0.00383}~&~\textcolor{blue}{0.00520}~\\
~&~\textcolor{green}{0.025}~&~\textcolor{green}{0.049}~&~\textcolor{green}{0.096}~&~\textcolor{green}{0.140}~\\
\hline
~10'~&~\textcolor{blue}{0.00159}~&~\textcolor{blue}{0.00248}~&~\textcolor{blue}{0.00385}~&~\textcolor{blue}{0.00522}~\\
~&~\textcolor{green}{0.018}~&~\textcolor{green}{0.039}~&~\textcolor{green}{0.083}~&~\textcolor{green}{0.120}~\\
\hline
~20'~&~\textcolor{blue}{0.00170}~&~\textcolor{blue}{0.00255}~&~\textcolor{blue}{0.00391}~&~\textcolor{blue}{0.00527}~\\
~&~\textcolor{green}{0.018}~&~\textcolor{green}{0.041}~&~\textcolor{green}{0.037}~&~\textcolor{green}{0.050}~\\
\hline
~30'~&~\textcolor{blue}{0.00183}~&~\textcolor{blue}{0.00265}~&~\textcolor{blue}{0.00399}~&~\textcolor{blue}{0.00536}~\\
~&~\textcolor{green}{0.021}~&~\textcolor{green}{0.050}~&~\textcolor{green}{0.033}~&~\textcolor{green}{0.049}~\\
\hline
\end{tabular}
\caption{Tolerance levels for {\it differential gain} as a function of instrument sensitivity and beamwidth (set by 
requiring that the fractional error induced in the inferred $r$ (first) and $M_{\nu}$ (second) do not exceed 10\%). We assume here worst-case-scenario scanning strategy.}
\end{center}
\end{table*}

\begin{table*}
\begin{center}
\begin{tabular}[c]{|c|c|c|c|c|}
\hline
~~&~1~&~2~&~4~&~6~\\
~~&~[$\mu$K-arcmin]~&~[$\mu$K-arcmin]~&~[$\mu$K-arcmin]~&~[$\mu$K-arcmin]~\\
\hline
~3'~&~0.03747~&~0.08735~&~0.19147~&~0.29286~\\
~~&~0.52~&~1.00~&~0.69~&~0.75~\\
\hline
~5'~&~0.03826~&~0.08896~&~0.19248~&~0.29564~\\
~~&~0.41~&~1.10~&~0.76~&~0.79~\\
\hline
~10'~&~0.04057~&~0.09274~&~0.19964~&~0.30476~\\
~~&~0.29~&~0.62~&~1.60~&~1.10~\\
\hline
~20'~&~0.04853~&~0.10531~&~0.21992~&~0.33236~\\
~~&~0.31~&~0.59~&~1.30~&~2.20~\\
\hline
~30'~&~0.05876~&~0.12186~&~0.24634~&~0.36842~\\
~~&~0.38~&~0.70~&~1.50~&~2.40~\\
\hline
\end{tabular}
\caption{Tolerance levels for {\it differential pointing} as a function of instrument sensitivity and beamwidth (set by 
requiring that the fractional error induced in the inferred $r$ (first) and $M_{\nu}$ (second) do not exceed 10\%). 
We assume here worst-case-scenario scanning strategy.} 
\end{center}
\end{table*}

\begin{table*}
\begin{center}
\begin{tabular}[c]{|c|c|c|c|c|}
\hline
~~&~1~&~2~&~4~&~6~\\
~~&~[$\mu$K-arcmin]~&~[$\mu$K-arcmin]~&~[$\mu$K-arcmin]~&~[$\mu$K-arcmin]~\\
\hline
~3'~&~\textcolor{green}{0.11628}~&~\textcolor{green}{0.33176}~&~\textcolor{green}{0.86288}~&~\textcolor{green}{1.41819}~\\
~~&~\textcolor{green}{0.23}~&~\textcolor{green}{0.27}~&~\textcolor{green}{0.37}~&~\textcolor{green}{0.45}~\\
\hline
~5'~&~\textcolor{green}{0.04476}~&~\textcolor{green}{0.12708}~&~\textcolor{green}{0.32614}~&~\textcolor{green}{0.53941}~\\
~~&~\textcolor{green}{0.12}~&~\textcolor{green}{0.12}~&~\textcolor{green}{0.16}~&~\textcolor{green}{0.20}~\\
\hline
~10'~&~\textcolor{green}{0.01348}~&~\textcolor{green}{0.03700}~&~\textcolor{green}{0.09432}~&~\textcolor{green}{0.15470}~\\
~~&~\textcolor{green}{0.047}~&~\textcolor{green}{0.12}~&~\textcolor{green}{0.09}~&~\textcolor{green}{0.94}~\\
\hline
~20'~&~\textcolor{green}{0.00490}~&~\textcolor{blue}{0.01267}~&~\textcolor{blue}{0.03116}~&~\textcolor{green}{0.05015}~\\
~~&~\textcolor{green}{0.023}~&~\textcolor{green}{0.054}~&~\textcolor{green}{0.059}~&~\textcolor{green}{0.059}~\\
\hline
~30'~&~\textcolor{blue}{0.00308}~&~\textcolor{blue}{0.00760}~&~\textcolor{blue}{0.01795}~&~\textcolor{blue}{0.02857}~\\
~~&~\textcolor{green}{0.015}~&~\textcolor{green}{0.032}~&~\textcolor{green}{0.079}~&~\textcolor{green}{0.064}~\\
\hline
\end{tabular}
\caption{Tolerance levels for {\it differential beamwidth} as a function of instrument sensitivity and beamwidth (set by 
requiring that the fractional error induced in the inferred $r$ (first) and $M_{\nu}$ (second) do not exceed 10\%). We assume here worst-case-scenario scanning strategy.} 
\end{center}
\end{table*}

\begin{table*}
\begin{center}
\begin{tabular}[c]{|c|c|c|c|c|}
\hline
~~&~1~&~2~&~4~&~6~\\
~~&~[$\mu$]K-arcmin~&~[$\mu$K-arcmin]~&~[$\mu$K-arcmin]~&~[$\mu$K-arcmin]~\\
\hline
~3'~&~\textcolor{green}{0.23213}~&~\textcolor{green}{0.65811}~&~\textcolor{green}{1.68434}~&~\textcolor{green}{2.72548}~\\
~~&~\textcolor{green}{0.76}~&~\textcolor{green}{1.5}~&~\textcolor{green}{3.0}~&~\textcolor{green}{4.4}~\\
\hline
~5'~&~\textcolor{green}{0.08964}~&~\textcolor{green}{0.25226}~&~\textcolor{green}{0.63814}~&~\textcolor{green}{1.03733}~\\
~~&~\textcolor{green}{0.28}~&~\textcolor{green}{0.58}~&~\textcolor{green}{1.2}~&~\textcolor{green}{1.8}~\\
\hline
~10'~&~\textcolor{green}{0.02693}~&~\textcolor{blue}{0.07404}~&~\textcolor{green}{0.18571}~&~\textcolor{green}{0.29928}~\\
~~&~\textcolor{green}{0.086}~&~\textcolor{green}{0.18}~&~\textcolor{green}{0.36}~&~\textcolor{green}{0.54}~\\
\hline
~20'~&~\textcolor{green}{0.00974}~&~\textcolor{green}{0.02526}~&~\textcolor{green}{0.06187}~&~\textcolor{green}{0.09868}~\\
~~&~\textcolor{green}{0.044}~&~\textcolor{green}{0.087}~&~\textcolor{green}{0.17}~&~\textcolor{green}{0.25}~\\
\hline
~30'~&~\textcolor{blue}{0.00605}~&~\textcolor{blue}{0.01512}~&~\textcolor{blue}{0.03588}~&~\textcolor{blue}{0.05665}~\\
~~&~\textcolor{green}{0.029}~&~\textcolor{green}{~0.054}~&~\textcolor{green}{0.10}~&~\textcolor{green}{0.14}~\\
\hline
\end{tabular}
\caption{Tolerance levels for {\it differential ellipticity} as a function of instrument sensitivity and beamwidth (set by requiring that the fractional error induced in the inferred $r$ (first) and $M_{\nu}$ (second) do not exceed 10\%).} 
\end{center}
\end{table*}

\begin{table*}
\begin{center}
\begin{tabular}[c]{|c|c|c|c|c|}
\hline
~~&~1~&~2~&~4~&~6~\\
~~&~[$\mu$K-arcmin]~&~[$\mu$K-arcmin]~&~[$\mu$K-arcmin]~&~[$\mu$K-arcmin]~\\
\hline
~3'~&~0.01982~&~0.04432~&~0.09047~&~0.13569~\\
~~&~0.23~&~0.38~&~0.62~&~0.84~\\
\hline
~5'~&~0.02040~&~0.04514~&~0.09100~&~0.13695~\\
~~&~0.24~&~0.34~&~0.58~&~0.79~\\
\hline
~10'~&~0.02232~&~0.04754~&~0.09443~&~0.14102~\\
~~&~0.42~&~0.45~&~0.65~&~0.87~\\
\hline
~20'~&~0.02867~&~0.05423~&~0.10274~&~0.15165~\\
~~&~0.54~&~0.60~&~0.88~&~1.20~\\
\hline
~30'~&~0.03492~&~0.06038~&~0.11119~&~0.16264~\\
~~&~0.46~&~0.70~&~1.10~&~1.70~\\
\hline
\end{tabular}
\caption{Tolerance levels for {\it beam rotation} as a function of instrument sensitivity and beamwidth (set by 
requiring that the fractional error induced in the inferred $r$ (first) and $M_{\nu}$ (second) do not exceed 10\%).}
\end{center}
\end{table*}

\newpage
\section{Discussion}

Gravitational lensing imprints the large-scale potentials along the line-of-sight to last scattering on the observed CMB and generates a guaranteed B-mode signal.
The most powerful technique for using this extra information is the quadratic estimator formalism, which extracts the lensing signal in the form of a noisy map $\hphi_{\ell m}$
of the lensing potential.
For the sky coverage, noise levels, and beam size expected for CMBpol, this indirect measurement of $\phi_{\ell m}$ will have high signal-to-noise on a wide range of
angular scales, and can be a source of cosmological information which is complementary to the primary CMB.
It is possible to place constraints on ``late universe'' parameters such as neutrino mass (Fig.~\ref{fig:neutrino_mass_forecast}), the dark energy equation of state
(Fig.~\ref{fig:w_forecast}), and curvature (Fig.~\ref{fig:curvature_forecast}) from the CMB alone.

In addition to being a source of cosmological information, gravitational lensing is also a contaminant of the gravitational wave B-mode signal on large scales.
The experimental requirement for lensing to be a limiting source of uncertainty is quite stringent: the instrumental
sensitivity must be $\sim 5$ $\mu$K-arcmin or better, and contamination from foregrounds and instrumental systematics must also be controlled
to better than this level.  However, if these requirements can be met, CMB experiments will enter a regime in which large-scale B-mode
measurements are intimately connected with small-scale measurements of the lensing B-mode: the only possibility for further improvement 
in $\sigma(T/S)$ will be to use ``delensing'' techniques which use the B-modes on small scales to reduce the level of lensing contamination.
The improvement in $\sigma(T/S)$ from delensing will depend on the instrumental noise level and beam.
Since both instrumental sensitivity and beam width are primary drivers of the total cost and complexity of a mission, weighing the tradeoffs
is likely to be a complex question when designing experiments, particularly since foregrounds and instrumental systematics will also be considerations.
Our main result (Fig.~(\ref{fig:delensing_polarization})) shows the dependence of the residual B-mode noise level on large scales, as a function of the noise and beam,
to help in this design decision.
``External'' delensing of the gravity wave B-mode, either via lens reconstruction from small-scale CMB {\em temperature} anisotropies or large-scale
structure, is not a promising approach; we establish ``no-go'' theorems (\S\ref{ssec:temperature_delensing}, \S\ref{ssec:lss_delensing})
showing that the improvement in $\sigma(T/S)$ is minimal, even under optimistic simplifying assumptions.

We have done a detailed analysis of the contamination to the lens reconstruction expected from polarized foregrounds, and concluded (\S\ref{sec:foregrounds})
that currently favored foreground models do not predict that residual foregrounds will be a significant source of bias.
Our $r$ and $M_{\nu}$ forecast in the presence of beam systematics illustrates (\S\ref{sec:systematics}) that the five types of systematics 
considered here will not significantly bias either the B-mode constraint on $(T/S)$ or the lens reconstruction constraint on neutrino mass.
We conclude that neither foregrounds nor beam systematics are likely to be a limiting factor to the promising science that lies ahead, as future
generations of experiments probe gravitational lensing through the CMB polarization on small angular scales.

\section*{Acknowledgements}

This work was organized and initiated at the workshop ``CMB Polarization workshop: theory and foregrounds'' held at Fermilab from Jun 23--26 2008.
We would like to thank the organizers and staff for a stimulating and productive atmosphere.

This work was supported by
an STFC Postdoctoral Fellowship (KMS),
NASA grant NNX08AU21G (AC),
NASA grant NNX08AH30G (SD),
NSF PHY-0555689 (MK),
NSF CAREER/PECASE Award AST-0548262 (BK),
and 
the DOE under DE-FG02-92ER40699 and the Initiatives in Science and Engineering Program at Columbia University (ML).

\newpage

\bibliographystyle{h-physrev3}
\bibliography{cmbpol_lensing}

\appendix

\newpage
\section{Methodology: lens reconstruction and delensing}
\label{app:lens_reconstruction}

\subsection{Lens reconstruction from CMB temperature and polarization}

We will construct the lens reconstruction estimator $\hphi_{\ell m}$ and its noise power spectrum in a uniform way which applies to both temperature
(used in \S\ref{ssec:temperature_delensing}) and polarization.  The notation in this appendix follows \cite{DvorkinSmith}.

In the presence of a nonzero lensing potential, the CMB two-point function acquires off-diagonal (i.e. $\ell\ne\ell'$, $m\ne m'$) correlations.
To lowest order in $\phi$, these take the form
\be
\langle a^X_{\ell_1m_1} a^Y_{\ell_2m_2} \rangle = \sum_{\ell m} \Gamma^{XY}_{\ell_1\ell_2\ell} \threej{\ell_1}{\ell_2}{\ell}{m_1}{m_2}{m} \phi^*_{\ell m}  \label{eq:gamma_2pt}
\ee
where $X,Y\in\{T,E,B\}$.  
(The two-point function in Eq.~(\ref{eq:gamma_2pt}) is the most general form which is linear in $\phi$ and satisfies overall rotation invariance.)
The $\Gamma$ couplings are given by:
\ba
\Gamma^{TT}_{\ell_1\ell_2\ell_3} &=& C_{\ell_1}^{TT} F^0_{\ell_2\ell_1\ell_3} + C_{\ell_2}^{TT} F^0_{\ell_1\ell_2\ell_3} \\
\Gamma^{TE}_{\ell_1\ell_2\ell_3} &=& C_{\ell_1}^{TE} \left( \frac{F^{-2}_{\ell_2\ell_1\ell_3} + F^{2}_{\ell_2\ell_1\ell_3}}{2} \right) + C_{\ell_2}^{TE} F^0_{\ell_1\ell_2\ell_3} \\
\Gamma^{EE}_{\ell_1\ell_2\ell_3} &=& C_{\ell_1}^{EE} \left( \frac{F^{-2}_{\ell_2\ell_1\ell_3} + F^{2}_{\ell_2\ell_1\ell_3}}{2} \right) +
                                     C_{\ell_2}^{EE} \left( \frac{F^{-2}_{\ell_1\ell_2\ell_3} + F^{2}_{\ell_1\ell_2\ell_3}}{2} \right)   \\
\Gamma^{TB}_{\ell_1\ell_2\ell_3} &=& C_{\ell_1}^{TE} \left( \frac{F^{-2}_{\ell_2\ell_1\ell_3} - F^{2}_{\ell_2\ell_1\ell_3}}{2i} \right)   \\
\Gamma^{EB}_{\ell_1\ell_2\ell_3} &=& C_{\ell_1}^{EE} \left( \frac{F^{-2}_{\ell_2\ell_1\ell_3} - F^{2}_{\ell_2\ell_1\ell_3}}{2i} \right)  \label{eq:gamma_eb_def}
\ea
where the $F$ symbol is defined by:
\be
F^s_{\ell_1\ell_2\ell_3} = [-\ell_1(\ell_1+1) + \ell_2(\ell_2+1) + \ell_3(\ell_3+1)]
  \sqrt{\frac{(2\ell_1+1)(2\ell_2+1)(2\ell_3+1)}{16\pi}} \threej{\ell_1}{\ell_2}{\ell_3}{-s}{s}{0}
\ee
The estimator $\hphi_{\ell m}$ is constructed as follows.
Assume signal + noise power spectra
\be
C_\ell + N_\ell = \left( \begin{array}{ccc}
 C_\ell^{TT} + N_\ell^{TT} & C_\ell^{TE} & 0  \\
 C_\ell^{TE} & C_\ell^{EE} + N_\ell^{EE} & 0 \\
 0 & 0 & C_\ell^{BB} + N_\ell^{BB}
\end{array} \right)
\ee
The minimum variance unbiased estimator and its noise power spectrum are given by:
\ba
N_\ell^{\phi\phi} &=& \left[ \frac{1}{2(2\ell+1)} \sum_{XYX'Y'\ell_1\ell_2}
\Gamma^{XY}_{\ell_1\ell_2\ell} (C_{\ell_1} + N_{\ell_1})^{-1}_{XX'}
\Gamma^{X'Y'*}_{\ell_1\ell_2\ell} (C_{\ell_2} + N_{\ell_2})^{-1}_{YY'}
\right]^{-1}  \label{eq:appendix_nlphi}  \\
\hphi_{\ell m} &=& \frac{N_\ell^{\phi\phi}}{2} \sum_{XY\ell_1m_1\ell_2m_2} \Gamma^{XY}_{\ell_1\ell_2\ell} \threej{\ell_1}{\ell_2}{\ell}{m_1}{m_2}{m} (C^{-1}a)^{X*}_{\ell_1m_1} (C^{-1}a)^{Y*}_{\ell_2m_2}
\ea
(Here and elsewhere in this appendix, we give expressions in harmonic-space form, but we note that for practical evaluation it is necessary to
rewrite them in a computationally efficient position-space form, see e.g. \cite{Okamoto:2003zw}.)

This construction assumes that both temperature and polarization are combined into a single minimum-variance estimator $\hphi_{\ell m}$, but
temperature-only and polarization-only reconstructions are the special cases $\{N_\ell^{EE},N_\ell^{BB}\}\rightarrow 0$ and $N_\ell^{TT}\rightarrow 0$
respectively.

\subsection{Delensing}

In the gradient approximation, the lensed B-mode is given in terms of the unlensed E-mode and the lens potential by:
\be
a^B_{\ell_2 m_2} = \sum_{\ell_1m_1\ell m} \Gamma^{EB}_{\ell_1\ell_2\ell} a^{E*}_{\ell_1m_1} a^{\phi *}_{\ell m} \label{eq:bmode_in_gradient_approx}
\ee
where $\Gamma^{EB}_{\ell_1\ell_2\ell}$ was defined in Eq.~(\ref{eq:gamma_eb_def}).  In this notation, the lensed B-mode power spectrum is given by \cite{Hu:2000ee}:
\be
C_{\ell_2}^{BB} = \frac{1}{2\ell_2+1} \sum_{\ell_1\ell} |\Gamma^{EB}_{\ell_1\ell_2\ell}|^2 C_{\ell_1}^{EE} C_\ell^{\phi\phi}
\ee
To perform delensing, we construct an estimator $\ha^B_{\ell m}$ for the lensing B-mode, given the reconstruction $\hphi_{\ell m}$
described in the previous subsection, and a (noisy) observation of the E-mode $\ha^E_{\ell m}$.  Heuristically, the estimator is
constructed by simply substituting the Wiener-filtered $\hE_{\ell m}$ and $\hphi_{\ell m}$ into Eq.~(\ref{eq:bmode_in_gradient_approx}):
\be
\ha^B_{\ell_2 m_2} = \sum_{\ell_1m_1\ell m} \Gamma^{EB}_{\ell_1\ell_2\ell} 
                      \frac{C^{EE}_{\ell_1} a^{E*}_{\ell_1m_1}}{C^{EE}_{\ell_1}+N^{EE}_{\ell_1}}
                      \frac{C^{\phi\phi}_{\ell} \hphi^*_{\ell m}}{C^{\phi\phi}_{\ell}+N^{\phi\phi}_{\ell}}  \label{eq:hblm_def}
\ee
The delensed B-mode power spectrum, i.e. the power spectrum of the residual field $(a^B_{\ell m}-\ha^B_{\ell m})$, is given by:
\be
C_{\ell_2}^{BB}\del =  \frac{1}{2\ell_2+1} \sum_{\ell_1\ell} |\Gamma^{EB}_{\ell_1\ell_2\ell}|^2 
\left(
C_{\ell_1}^{EE} C_\ell^{\phi\phi} -
\frac{(C_{\ell_1}^{EE} C_\ell^{\phi\phi})^2}{(C_{\ell_1}^{EE}+N_{\ell_1}^{EE})(C_\ell^{\phi\phi}+N_\ell^{\phi\phi})}
\right)  \label{eq:clbb_del}
\ee
(A more formal derivation of the estimator in Eq.~(\ref{eq:hblm_def}) can be obtained by solving for the weights in the estimator which minimize
the delensed B-mode power spectrum.)

The noise power spectrum $N_\ell^{\phi\phi}$ in Eq.~(\ref{eq:appendix_nlphi}) and the residual B-mode power spectrum $C_\ell^{BB}\del$ in Eq.~(\ref{eq:clbb_del})
were obtained assuming quadratic lens reconstruction.
If we want to assume iterative lens reconstruction \cite{Hirata:2002jy,Hirata:2003ka}, then we simply repeat the calculation
of $N_\ell^{\phi\phi}, C_\ell^{BB}\del$ replacing the lensed B-mode power spectrum (which enters the calculation of $N_\ell^{\phi\phi}$ as a source
of noise in Eq.~(\ref{eq:appendix_nlphi})) by the current value of $C_\ell^{BB}\del$, and iterate until $C_\ell^{BB}\del$ converges.

We calculate $N_\ell^{\phi\phi}, C_\ell^{BB}\del$ using the scheme described above when making forecasts throughout this report.
It is important to note that this scheme is actually an approximate forecasting procedure for computing these power spectra and is not exact.
For example, a complete calculation of the power spectrum $C_\ell^{BB}\del$ to lowest order in $\phi$ would include an 8-point correlation
function containing 6 factors of $E$ and 2 factors of $\phi$.
In this language, the approximate result in Eq.~(\ref{eq:clbb_del}) would equal a subset of all the contractions obtained using Wick's theorem.
To show that our approximate scheme is in fact a good approximation, we compared forecasts obtained using this scheme with the
Monte Carlo simulations of delensing in \cite{Seljak:2003pn} for specific choices of noise level and beam, and find good agreement.

Finally, we discuss forecasts for the parameter uncertainty $\sigma(T/S)$.
As mentioned in \S\ref{sec:delensing}, the value of $\sigma(T/S)$ achievable in a given experiment will depend not only on the instrumental noise and beam,
but also on sky coverage and loss of modes at low $\ell$ due to EB mixing from survey boundaries, or from projecting out foregrounds.
To quantify this latter set of complications, we introduce a mode density function $(dN_{\rm modes}/d\ell)$ to represent the number of modes that
can be measured at a given value of $\ell$, and write the parameter uncertainty $\sigma(T/S)$ with and without delensing as:
\ba
\sigma(T/S)_{\rm no\ delensing} &=& \left[ \frac{1}{2} \sum_\ell \left( \frac{dN_{\rm modes}}{d\ell} \right) 
  \left( \frac{C_\ell^{BB}({\rm tensor})}{C_\ell^{BB}({\rm lensed}) + (\sigma^B_{\rm inst})^2} \right)^2 \right]^{-1/2} \label{eq:appendix_ts1} \\
\sigma(T/S)_{\rm with\ delensing} &=& \left[ \frac{1}{2} \sum_\ell \left( \frac{dN_{\rm modes}}{d\ell} \right) 
  \left( \frac{C_\ell^{BB}({\rm tensor})}{C_\ell^{BB}\del + (\sigma^B_{\rm inst})^2} \right)^2 \right]^{-1/2}  \label{eq:appendix_ts2}
\ea
(We have normalized $dN_{\rm modes}/d\ell$ so that $dN_{\rm modes}/d\ell = (2\ell+1)$ for an all-sky survey.)

Empirically, we find that both the lensed and delensed B-mode power spectra are constant at low $\ell$ to an excellent approximation:
\ba
C_\ell^{BB}({\rm lensed}) &\approx& (\sigma^B_{\rm lensed})^2  \qquad (\ell\le 100) \\
C_\ell^{BB}({\rm delensed}) &\approx& (\sigma^B_{\rm delensed})^2  \qquad (\ell\le 100)
\ea
(To quantify this better, the difference between each power spectrum and its best-fit constant approximation is below the cosmic variance limit
for detectability using only multipoles with $\ell\le 100$.)
From this and Eqs.~(\ref{eq:appendix_ts1}),~(\ref{eq:appendix_ts2}) it follows that:
\be
\frac{\sigma(T/S)_{\rm no\ delensing}}{\sigma(T/S)_{\rm with\ delensing}} = 
\frac{(\sigma^B_{\rm lensed})^2 + (\sigma^B_{\rm inst})^2}{(\sigma^B_{\rm delensed})^2 + (\sigma^B_{\rm inst})^2}
\ee
We have used this simplication throughout \S\ref{sec:delensing}.

\newpage
\section{Methodology: Fisher forecasts}
\label{app:fisher}
As discussed in the text, CMB lensing reconstruction allows us to add an additional source of information to the usual CMB temperature and polarization fields  --- namely, the lens reconstructed deflection field, $d_l^m$.  The deflection field contains information about late time geometry and structure in the universe and helps  break the angular diameter distance degeneracy in the CMB. \par

For the purpose of Fisher Matrix calculations, it is  profitable to assume that the lens reconstruction has been used to \emph{de-lens} the temperature and polarization fields, yielding four Gaussian independent variables $\{T_l^m, E_l^m, B_l^m, d_l^m\}$, which are the unlensed CMB fields plus the deflection modes.  If we further assume, as we have done here, that the fiducial model has no primordial $B$ mode, then the de-lensed $B$-mode is purely noise. Omitting the latter from the Fisher calculation, the data covariance matrix reads, 
\begin{equation}
\bf C_\ell =\left (\begin{array}{ccc} 
C_\ell^{TT}+N_\ell^{TT} &C_\ell^{TE} &  C_\ell^{Td}\\
C_\ell^{TE} & C_\ell^{EE}+N_\ell^{EE} & 0 \\
 C_\ell^{Td } &  0  & C_\ell^{dd}+N_\ell^{dd}
\end{array} \right )
\end{equation}
where the $C_\ell^{XY}$ 's  are the unlensed power spectra and the $N_\ell^{XX}$'s denote noise power spectra.  The deflection field power spectrum $N_\ell^{dd}$ can be computed  in the context of a quadratic estimator for the deflection field, as prescribed in \cite{Okamoto:2003zw}. \par
Under these assumptions, the Fisher Matrix can be simply written as, 
\begin{equation}
F_{ij} = \sum_\ell \frac{(2\ell+1)}{2}f_{\mathrm{sky}}  \mathrm{Trace}\left[\mathbf C_\ell ^{-1}\frac{ \partial \mathbf C_\ell}{\partial p_i}   C_\ell ^{-1}\frac{ \partial \mathbf C_\ell}{\partial p_j}\right]
\end{equation}
where $p_i$ denotes the $i$-th cosmological parameter and the lower bound on the error on $p_i$ after marginalization over all other free parameters is given by,
\begin{equation}
\sigma{(p_i)}=\sqrt{(F^{-1})_{ii}}.
\end{equation}
For the cases considered in this report we considered a standard $6$ parameter $\Lambda$CDM cosmology, parameterized via $\{\Omega_b h^2,\Omega_{\mathrm{DM}} h^2,\theta_A, \tau,n_s,A_s\}$ and extended it to include a massive neutrino density as a fraction $f_\nu$ of the total dark matter density,  a dark energy equation of state $w$, and a curvature energy density $\Omega_k$.  Note that we chose $\theta_A$, the angular scale of the sound horizon at recombination, as a parameter, rather than $\Omega_\Lambda$. This is crucial because $\theta_A$ is the observed quantity and should be kept constant when evaluating the derivatives with respect to other parameters. We ensured  the convergence of the Fisher calculations by repeating all exercises after halving  the step sizes and making sure that the derivatives and constraints on parameters remain effectively  unchanged.  We chose a fiducial  model given by $\{\Omega_b h^2,\Omega_{\mathrm{DM}} h^2,\theta_A, \tau,n_s,(10^9~A_s),f_\nu,w,\Omega_k\}= \{0.023,0.121,0.010464,0.11, 0.96,2.453, 0.008,-1,0\}$ with one massive neutrino species. 

\end{document}